\documentclass[twocolumn,pre,superscriptaddress]{revtex4}
\providecommand{\JournalTitle}[1]{#1}
\usepackage{amssymb, bm, amsmath}
\usepackage{xcolor, graphicx}
\usepackage[T1]{fontenc}

\graphicspath{{./figures/}}

\DeclareMathOperator*{\argmax}{arg\,max}

\newcommand{\beq}{\begin{equation}}
\newcommand{\eeq}{\end{equation}}
\newcommand{\quotes}[1]{``#1''}
\newcommand{\lastequal}{Corresponding authors. These authors contributed equally.}

\newcommand{\SMRef}[1]{{{#1}}}

\newcommand{\rev}[1]{{#1}}
	
\renewcommand{\textendash}{\text{--}}

\begin{document}
		
\title{Dynamics of memory B cells and plasmablasts in healthy individuals}
		
\newcommand{\ENS}{{Laboratoire de physique de l'\'Ecole normale sup\'erieure,
		CNRS, PSL University, Sorbonne Universit\'e, and Universit\'e 
		Paris-Cit\'e, 75005 Paris, France}}
	
\author{Andrea Mazzolini}
\affiliation{\ENS}
\affiliation{Department of Physics and INFN, University of Turin, 10125 Turin, Italy}
\affiliation{INFN, Sezione di Bari, Italy}
\author{Aleksandra M. Walczak}
\thanks{\lastequal}
\affiliation{\ENS}
\author{Thierry Mora}
\thanks{\lastequal}
\affiliation{\ENS}
		
\begin{abstract}
Our adaptive immune system relies on the persistence over long times of a diverse set of antigen-experienced B cells to encode our memories of past infections and to protect us against future ones. While longitudinal repertoire sequencing promises to track the long-term dynamics of many B cell clones simultaneously, sampling and experimental noise make it hard to draw reliable quantitative conclusions. Leveraging statistical inference, we infer the dynamics of memory B cell clonal dynamics and conversion to plasmablasts, which includes clone creation, degradation, abundance fluctuations, and differentiation. We find that memory B cell clones degrade slowly, with a half-life of 10 years. Based on the inferred parameters, we predict that it takes about 50 years to renew 50\% of the repertoire, with most observed clones surviving for a lifetime. We infer that, on average, 1 out of 100 memory B cells differentiates into a plasmablast each year, more than expected from purely antigen-stimulated differentiation, and that plasmablast clones degrade with a half-life of about one year in the absence of memory imports. Our method is general and could be applied to other longitudinal repertoire sequencing B cell subsets.

\end{abstract}
	
\maketitle

\section{Introduction}

The adaptive immune system has to recognize a practically infinite reservoir of antigens derived from pathogens.
This is achieved thanks to the huge diversity of lymphocyte receptors created through V(D)J recombination, which randomly generates new sequences \cite{tonegawa1983somatic, alt1992vdj}, and can lead to $10^8-10^{10}$ distinct receptor types in each organism \cite{lythe2016many, altan2020quantitative}.
This diversity is further increased in B cells through affinity maturation  \cite{maclennan1994germinal, victora2022germinal}, after rounds of somatic hypermutations and selection against encountered antigens \cite{elhanati2015inferring, hoehn2016diversity}.
This process releases cells in the blood that are evolutionarily related, forming a clonal family specifically tuned for the antigen against which it evolved.
These cells divide into memory cells and plasma cells \cite{shlomchik2012germinal} whose role is to keep immunological memory against encountered antigens.
In particular, memory cells can differentiate into antibody-secreting plasmablasts upon antigen re-encounter \cite{lam2024guide}, providing a quick and effective response to reinfections.

High-throughput DNA-sequencing experiments have significantly increased our knowledge about the immune repertoire diversity \cite{boyd2009measurement, six2013past, elhanati2015inferring, dewitt2016public, soto2019high, briney2019commonality} and evolutionary properties of B-cell evolution \cite{mccoy2015quantifying, horns2019signatures, olson2018bayesian, ralph2020using, spisak2020learning}.
Furthermore, sequencing experiments taken at multiple time points opened up several questions related to B-cell repertoire dynamics \cite{horns2019signatures, nourmohammad2019fierce, hoehn2021human, mazzolini2022inspecting} and, in particular, about the stability of the memory cell repertoire and its relation with plasmablasts in healthy conditions \cite{phad2022clonal, mikelov2022memory}.
The memory repertoire was found to be  relatively stable at the level of comparing statistics of common clones between time points and across experimental replicates at the same time. 
The authors in \cite{phad2022clonal} hypothesized that memory-to-plasmablast differentiation does not happen only upon specific antigen re-encounter \cite{ochsenbein2000protective}, but also through cross-reactivity with different persisting antigens or antigen-independent mechanisms \cite{bernasconi2002maintenance, horns2020memory}. However, the timescales of these events remain unknown.

Learning about the dynamics of repertoires from data requires a lot of care. Limited sampling and noise in repertoire sequencing experiments leads to low clonal overlap between samples, including between technical replicates taken at the same timepoint \cite{phad2022clonal}. Such observations in longitudinal studies may wrongly suggest a large turnover. In this work, we aim to provide a quantitative description of \rev{the repertoires of two cell types: memory B cells and plasmablasts.
To this end, we apply} statistical inference techniques to sequencing data from the blood of healthy individuals.
To disentangle the actual repertoire dynamics from the large sampling and experimental noise that confounds it, we adapt a previously introduced method used in the context of T-cell repertoire \cite{puelma2020inferring, koraichi2022noiset, bensouda2023inferring}.
Our inference provides biologically informative, previously unknown quantities, such as the half-life of memory clones, how quickly the repertoire replaces its B-cell clones, or the average differentiation rate of memory cells into plasmablasts. 
The model accurately describes the statistics of clone abundances across longitudinal experiments, providing probabilistic predictions of future cell counts given present observations in healthy conditions. This can provide a useful baseline for tracking the evolution of immunity and response to chronic and recurring pathologies over time.

\section{Results}

\begin{figure*}
	\centering
	\includegraphics[width=1\linewidth]{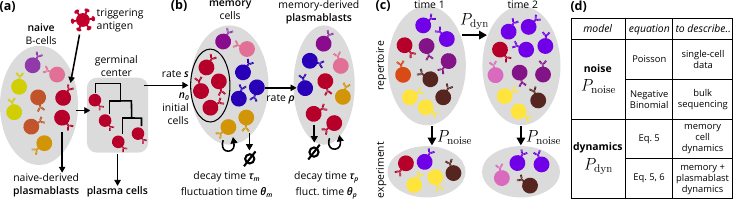}
	\caption{\textbf{Modeling and inference scheme.}
		(a): Simplified schematic of biological processes that lead to the creation of memory-cell clones starting from the naive repertoire.
		(b): Processes involving memory and plasmablast clones that are included in the studied dynamical models (see Sec. \ref{sec:scheme}).
		\rev{Empty sets represent cell death and
                  self-arrows represent proliferation.}
		(c): Inference scheme of the repertoire clone dynamics between two time points. The variation of frequencies in the body is modeled with $P_{\rm{dyn}}$.
		Experiments provide counts that convey noisy estimates of the real abundances through the noise model $P_{\rm{noise}}$.
		(d): Different choices for the noise and the dynamical model discussed in the paper.
	}
	\label{fig:scheme}
\end{figure*}

\subsection{Scheme for inferring B cell population dynamics from noisy data}
\label{sec:scheme}

Our goal is to infer a probabilistic dynamical model, $P_{\rm dyn}$, to describe the evolution of the size of B cell clones over long times. We focus on two cell types: memory B cells and memory-derived plasmablasts. A summary of our model assumptions informed by their known biology is illustrated by Fig. \ref{fig:scheme}a,b. Memory B cell clones emerge from affinity maturation, with \rev{mean} rate $s$ and \rev{average} initial size $n_0$. They then follow stochastic dynamics subject to random environmental fluctuations. Mathematically, their clone sizes are described by a Geometric Brownian Motion (GBM), following \cite{desponds2016fluctuating, bensouda2023inferring}, see Methods. In that model, the size of each clone is subject to random fold-changes over a typical timescale $\theta_m$, \rev{which we refer to as the fluctuation time}. Overall, these changes, \rev{which can be caused by cell proliferation, death and differentiation}, tend to decrease the size of clones with characteristic time $\tau_m$, \rev{called decay time}. We chose that model because it is the simplest possible model for stochastic growth and decay, but also because it predicts power-law distributions of clone sizes \cite{desponds2016fluctuating}, consistent with experimental observations \cite{weinstein2009high, mora2018quantifying}. 
Within this framework, the exponent of the power law of the cumulative size distribution, $\alpha$, is linked to the dynamical parameters $\theta_m$ and $\tau_m$ through the relation $\alpha=2 \theta_m / \tau_m$.

Memory cells can differentiate into plasmablasts with rate $\rho$, providing a rapid response against pathogens encountered in previous infections.
After differentiation, plasmablast dynamics is also subject to environmental fluctuations and decay, which can be modeled by a GBM with \rev{timescales} $\theta_p$ and $\tau_p$, similarly to memory cells. The model we use to describe the dynamics of plasmablast repertoires combines the contributions of the GBM with a source of new cells from memory B cell differentiation (see Methods).

\rev{
The biological processes we aim to describe are discrete, heterogeneous in time and across clones---for example, infections cause a small subset of clones to undergo quick antigen-driven bursts. The applicability and interpretation of our model, which summarizes these heterogeneities into stochastic terms, make sense only when looking at samples that are well separated in time, on timescales larger than those typical events of a few weeks. Model rates represent a net average of the changes occurring over those long times. For example, $\rho$ should be interpreted as the fraction of memory cells that differentiate into plasmablasts in a year, rather than the rate of each individual cell.
}

To infer those dynamical models, we use longitudinal repertoire sequencing data. 
Repertoire sequencing only probes an infinitesimal fraction of the immune repertoire, yielding a noisy estimate of clonal frequencies.  At the same time, the repertoire dynamics themselves are expected to be stochastic and modeled by $P_{\rm dyn}$, following, for example, the GBM described above. To disentangle these two sources of stochasticity --- experimental noise and biological fluctuations --- we developed a statistical inference scheme based on  \cite{puelma2020inferring, bensouda2023inferring, koraichi2022noiset} and presented in the Methods section, with additional detail in the \SMRef{SM}. It relies on two probabilistic models  (Fig. \ref{fig:scheme}c): the dynamical model $P_{\rm dyn}$ and a noise model $P_{\rm noise}$, which describes how actual clone sizes translate into observed sequence counts. It is chosen to be a simple Poisson sampling model or a more complex Negative Binomial model, depending on the nature of the data (more on this below). The different types of dynamical and noise models used in this paper are summarized in Fig. \ref{fig:scheme}d.

\subsection{Longitudinal, replicated B cell repertoire datasets}

The method takes as input longitudinal datasets of memory B cells and circulating plasmablasts, and must also include technical replicates to calibrate the noise model.
We collected two open-access longitudinal datasets from human Peripheral Blood Mononuclear Cells (PBMC) that fit these requirements \cite{phad2022clonal}, \cite{mikelov2022memory}.

The first dataset \cite{phad2022clonal}, which we call {\em Dataset A}, comprises repertoires from two healthy individuals, aged 50 and 69, over a span of 10 and 6 years respectively, with up to 5 replicates per time point. All repertoires were obtained by single-cell sequencing, which allows for pairing heavy and light chains, as well as phenotyping. The samples, including their cell counts, are summarized in \SMRef{Fig. S1}.

The second dataset \cite{mikelov2022memory}, {\em Dataset B}, includes IGH (heavy-chain) repertoires from 6 healthy individuals, aged 23 to 39, spanning a period of 1 year, with up to 2 replicates per timepoint (see details and sequence counts in \SMRef{Fig. S2}). Cells were first sorted into memory B cells, plasma cells, and plasmablasts, and their repertoire was sequenced in bulk from mRNA, using Unique Molecular Identifiers to mitigate PCR noise and biases. We only used the repertoires of 3 of the 6 individuals (AT of age 23, IM of age 39, and IZ of age 33) for whom samples were sequenced deeply enough for all time points and cell types.

Data was processed using standard tools. We define a clone or clonal group as the set of B cells that descend from the same initial V(D)J recombination event. Clonal groups of sequences were inferred using HILARy \cite{spisak2024combining} (see \rev{Methods} for details).

\subsection{Distinct noise profiles of single-cell versus bulk repertoire sequencing}
\label{sec:noise}

\begin{figure}
	\centering
	\includegraphics[width=1\linewidth]{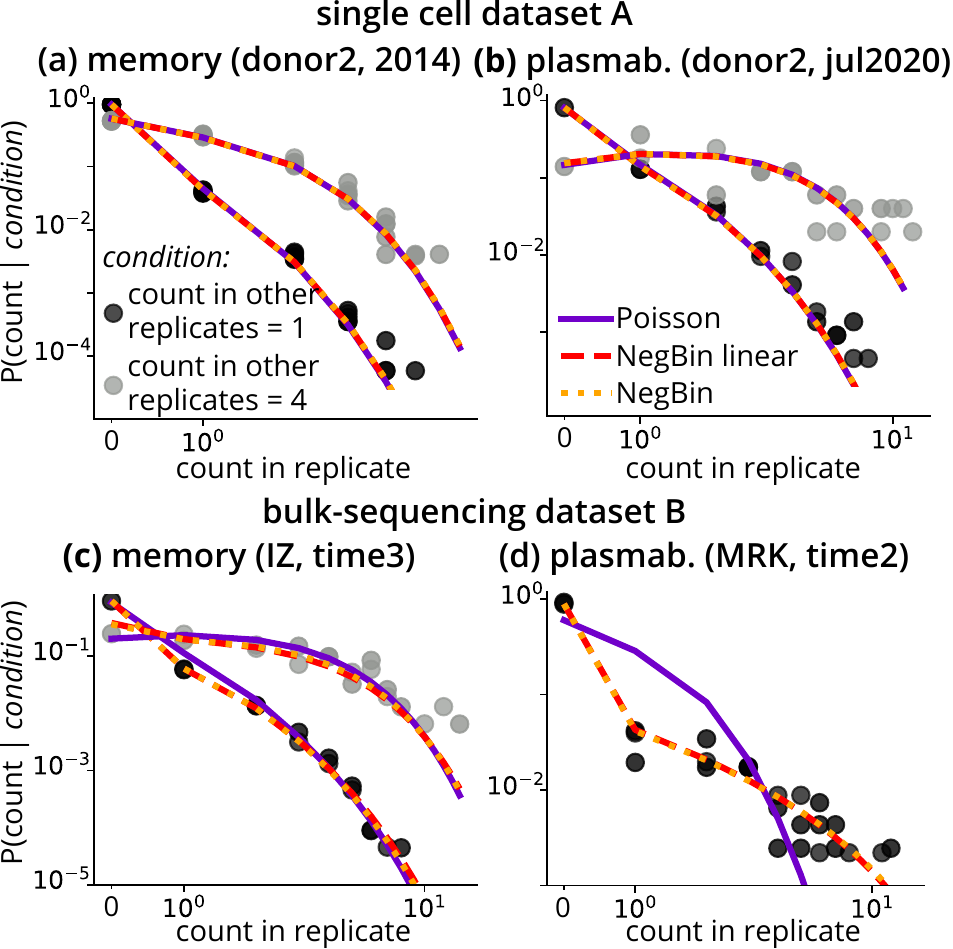}
	\caption{{\bf Validation of the noise model.} Shown is the conditional distribution of counts of a clone in one of the replicates, given the total count of the same clone in all other replicates (1 in black, 4 in gray). Each replicate is downsampled in such a way that every replicate has the same size to compare them with the same theoretical prediction (which depends on the sample size). Each panel shows a different experiment, corresponding to one donor in each of {\em Dataset A} (panels a and b), which uses single-cell sequencing, and {\em Dataset B}, which uses mRNA bulk sequencing (c and d), and to memory (a and c) as well as plasmablast (b and d) subsets. In (d) we do not report the conditioning on $4$ because the points partially overlap with the other case.
	For each sample, predictions from the inferred noise models are shown for the Poisson model, the best Negative Binomial Model with a linear relation between variance and mean (NegBin linear), and the best Negative Binomial Model with a power relation between variance and mean (NegBin).
	}
	\label{fig:noise_marginals}
\end{figure}

\rev{The first step of the inference scheme is to learn a model to quantify experimental and sampling noise ($P_{\rm noise}$ in Fig. \ref{fig:scheme}).}
The noise model comes in three versions of increasing complexity and generality (see Methods): the Poisson Model (zero parameters), a Negative Binomial Model with a linear relation between variance and mean (1 parameter), and a Negative Binomial Model with a power-law dependence of variance and mean (2 parameters). In addition, all 3 models assume that the distribution of clone frequencies $f$ follows a power law, parametrized by 2 additional parameters: its exponent $\alpha$ and the minimal clone frequency $f_{\rm min}$. The form of the clone size distribution is not strictly part of the noise model and is not expected to impact its inference, but it is necessary as a prior for the fitting procedure. We chose its form according to empirical observations.

To determine which model to use, we performed the inference of these three models on all the timepoints for which at least 2 replicates are available and compared their log-likelihood to assess their ability to accurately describe the data (\SMRef{Fig. S3 and S4}).
For the single-cell {\em Dataset A}, all three models performed equally well, leading us to choose the simpler Poisson Model for the rest of the analysis. 
This is in agreement with previous observations in single-cell RNAseq data \cite{pan2023poisson, lazzardi2023emergent}, consistent with a model in which the main source of noise is due to sampling single cells.
For {\em Dataset B}, which was obtained by bulk sequencing of mRNA, we found that the Negative Binomial Models performed slightly better than Poisson in the memory-cell samples and significantly better in plasmablasts. However, the more general model with a power-law dependence between variance and mean did not offer a substantial improvement over the model with a linear one. In addition, the power-law model was harder to optimize, finding parameters at the boundaries of the allowed space and violating prior knowledge, notably on the predicted power-law exponent of the clone size distribution (\SMRef{Fig. S4b}).
Therefore, for {\em Dataset B} we used the linear Negative Binomial Model, with a distinct linear coefficient inferred for each timepoint. The overdispersion of the measured count relative to mere sampling (Poisson) noise can be explained by the additional expression, reverse-transcription, and amplification noises caused by the mRNA bulk sequencing protocol \cite{puelma2020inferring, bensouda2023inferring, koraichi2022noiset}.

To test the noise model, we used it to predict the distribution of counts of a clone in one of the replicates, conditioned on its total count across the other replicates. Comparing this prediction to conditional histograms of clone sizes collected over the entire repertoire from a single sample yields excellent agreement (Fig. \ref{fig:noise_marginals}), except for the Poisson Model applied to {\em Dataset B}, especially for plasmablasts (Fig. \ref{fig:noise_marginals}d), consistent with our log-likelihood performance assessment.
Finally, the distribution of aggregated counts across all replicates follows a power law that is well predicted by the model, which validates the choice of the frequency distribution $\rho(f)$ (\SMRef{Fig. S5}).

\rev{
The inferred noise model should not depend on the size and number of used replicates. 
To check for this robustness, we repeated the noise inference on the 2014 memory sample of donor 2 of \emph{Dataset A} using various subsets of the 5 replicates (Fig. \SMRef{S3e}). The number of replicates mostly affected the inference of the power law exponent and minimal frequency, but only weakly, with a still negligible gap between the Negative Binomial and Poisson likelihoods, confirming our choice to use the Poisson model for this dataset.
}

\subsection{Memory B cell clones persist for decades}
\label{sec:memory_tau}

\begin{table*}[htbp]
	\begin{tabular}{c|c||c||c|c|c|c|c|c} 
	\multicolumn{2}{c|}{  } & \textbf{Noise model} & \multicolumn{6}{c}{ \textbf{Memory cell dynamics} } \\ [0.5ex] 
    & age & Neg. Bin. & decay time, & fluctuation & power law & initial clone & rate of new & number of \\
	& (year) & parameter, $a$ & $\tau_m$ (year) & time $\theta_m$ (year) & exponent, $\alpha$ & size, $n_0$ & clones, $s$ (1/year) & clones, $N_{\rm clone}$ \\ [0.5ex] 
	\hline 
	& & & & & & & & \\[-1.8ex]
	donor 1 (A) & 69 & 0 & $14.6 \pm 1.3$ & $9.3 \pm 0.6$ & $1.28 \pm 0.05$ & \rev{$4 \cdot 10^{4} \textendash 4 \cdot 10^5$} &  \rev{$(3.6 \pm 0.6) \cdot 10^3$} & $(5.5 \textendash 6.8) \cdot 10^5$ \\ 
	donor 2 (A) & 50 & 0 & $14.5 \pm 1.3$ & $\rev{8.7} \pm 0.7$ & $1.20 \pm 0.04$ & \rev{$1 \cdot 10^{4} \textendash 1 \cdot 10^5$} & \rev{$(9.0 \pm 1.6) \cdot 10^3$} & $(1.2 \textendash 1.5) \cdot 10^6$  \\
	AT (B) & 23 & \rev{$0.086 \pm 0.005$} & $> 3.2$ & $> 2.1$  &  \rev{$1.29 \pm 0.01$} & \rev{$5 \cdot 10^{3} \textendash 5 \cdot 10^4$}  & $< 1.5 \cdot 10^5$  & $(4.0 \textendash 5.1) \cdot 10^6$ \\
	IM (B) & 39 & \rev{$0.059 \pm 0.004$} & $> 4.0$ & $> 2.7$ & \rev{$1.30  \pm 0.01$} & \rev{$6 \cdot 10^{3} \textendash 6 \cdot 10^4$}  & $< 9.4 \cdot 10^4$  & $(3.4 \textendash 4.3) \cdot 10^6$ \\
	IZ (B) & 33 & \rev{$0.167 \pm 0.004$} & $> 4.8$ & $> 3.3$ & \rev{$1.35 \pm 0.01$} & \rev{$3 \cdot 10^{3} \textendash 3 \cdot 10^4$}  & $< 1.8 \cdot 10^5$  & $(6.8 \textendash 8.8) \cdot 10^6$ \\ [1ex] 
\end{tabular}
	\caption{
	\textbf{Inferred parameters of the Geometric Brownian Motion for memory B-cell dynamics.}
	The third column corresponds \rev{to the average parameter $a$ of the Negative Binomial model (see Methods) and its average error across the time points} (\rev{see Fig. S4a for values at each single time point}). The other columns give the GBM parameters ($\tau_m$, $\theta_m$, $s$, and $n_0$), and their derivatives, including the predicted exponent $\alpha=2 \theta / \tau$, and the repertoire diversity $N_{\rm clone}$.
	For $n_0$ and $N_{\rm clone}$ we show a range of \rev{values} corresponding to an assumed total number of memory B-cells \rev{$N_{\rm cell}=10^{10} \textendash 10^{11}$}. For \rev{details} about inference and error estimation see \SMRef{Sec. S3}.
}
\label{tab:params_gbm}
\end{table*}

\begin{figure}
	\centering
	\includegraphics[width=1\linewidth]{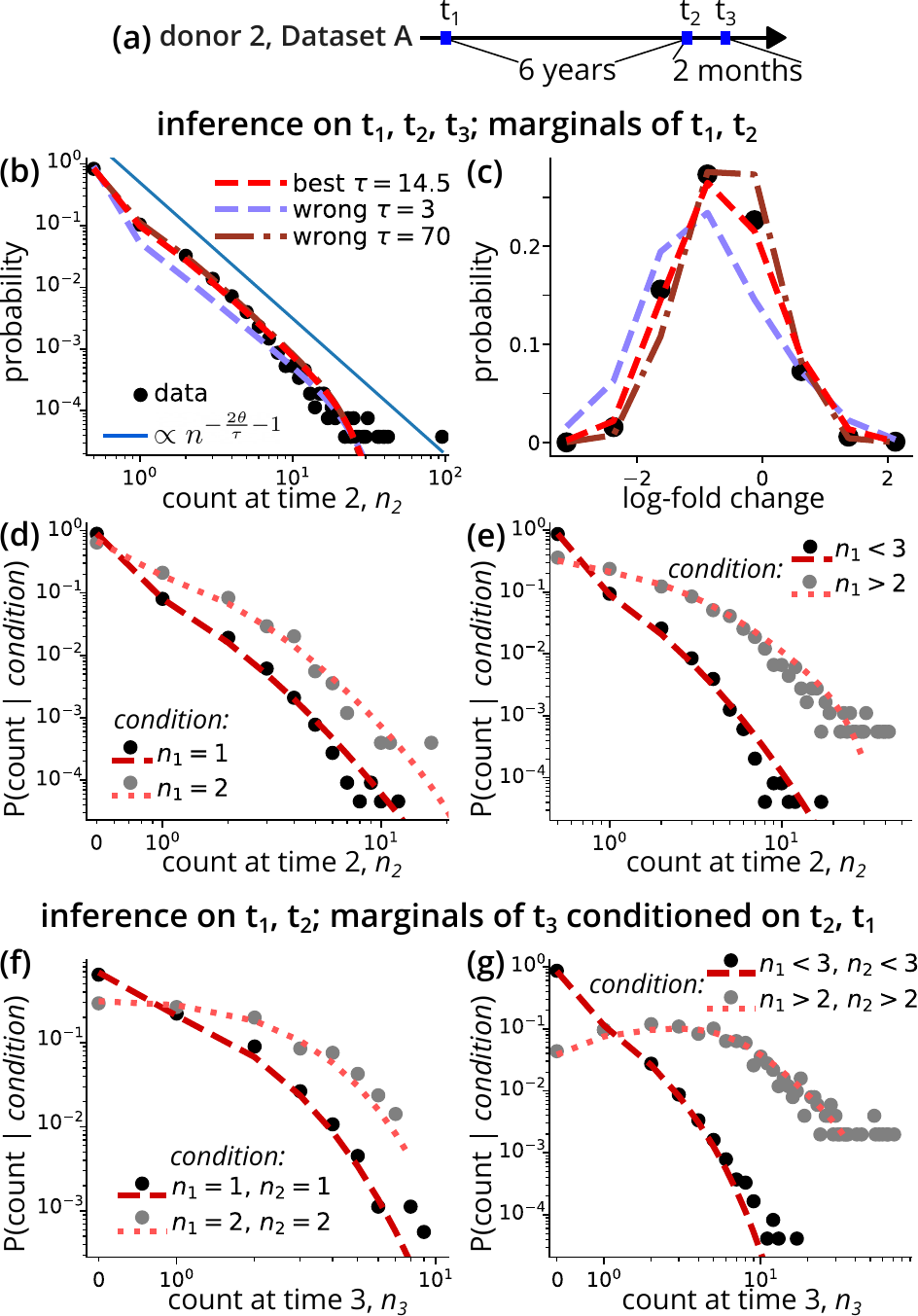}
	\caption{{\bf Validation of the Geometric Brownian Motion model for memory cells.}
	Statistics of memory B-cell counts and its prediction by the GBM model in donor 2 from {\em Dataset \rev{A}}.
    The model prediction is shown for the best fit value of $\tau_m=14.5$ yr and $\theta_m=10.1$ yr, as well as for $\tau_m=5$ and 70 yr, keeping the ratio $\theta_m/\tau_m$ constant to its best-fit value. 
	(a): Summary of the samples and timepoints of donor 2.
    (b): Clone size distribution at the second time point. 
	(c): Distribution of log-fold changes in clone counts between
        $t_1$ and $t_2$. \rev{We only include clones with count $n_1 > 2$.}
	(d), (e): Counts at time 2 given a condition on the counts observed at time 1. \rev{Black and gray points refer to two different conditions.}
	(f),(g): Similar to (d), (e) but for the third time point. To make a fully independent prediction, for these two panels the model parameters were re-learned using only the first two time points.
	}
	\label{fig:gbm_marginals}
      \end{figure}

We next applied our inference method to learn the dynamical parameters of the memory repertoire: the \rev{fluctuation time}, $\theta_m$, the net decay time, $\tau_m$, the rate of introduction of new memory clones, $s$, and their initial size, $n_0$. Parameter optimization was performed using Monte Carlo integration, as explained in detail in \SMRef{Sec.~S3} and illustrated in \SMRef{Fig.~S6}. The results are summarized in Table \ref{tab:params_gbm}.
The net decay time of individual clones in the memory repertoire of individuals in {\em Dataset A}  is $\approx 15$ years, or a half-life of $\ln(2)\times 15\approx 10$ years, meaning that, on average, clones reduce their size by half in 10 years. The noise timescale is $\approx 10$ years, meaning that it takes less than a decade for individual clone fluctuations to lead to a two-fold increase or decrease.

Inference on {\em Dataset B} does not allow us to learn a precise number for $\tau_m$, but puts on it a lower bound of 4 years (\SMRef{Sec. S3F and Fig. S7}), which is consistent with the estimate of 15 years from {\em Dataset A}.
The reason for that uncertainty is that the time interval between samples is only 1 year. Within that short interval, the data do not show evidence of a decrease in time, and the inference cannot distinguish between scenarios with a $\tau_m$ larger than the bound.

To assess the model's validity, we perform a similar test to the one performed for the noise model. We predict the distribution of clone counts, and of their fold change, as a function of their count at an earlier time point (instead of a replicate, as we did for the noise model). Despite its simplicity, the model fits \rev{all the distributions very well} (Fig. \ref{fig:gbm_marginals}b-g).
To show that the goodness of fit depends sensitively on learning the correct parameters, we also compared the prediction of sub-optimal models where the values of $\tau_m$ and $\theta_m$ were varied while keeping their ratio fixed to the inferred value. Theory predicts that the clone size distribution should follow approximately a power law, whose exponent depends only on that ratio, predicted to be $\alpha=2\theta_m/\tau_m\approx 1.2$-$1.3$. Consistently, that distribution is well captured even by the sub-optimal models (Fig. \ref{fig:gbm_marginals}b). By contrast, the distribution of log fold changes is only well reproduced by the inferred model ($\tau_m=14.5$ yr), and not by the sub-optimal ones ($\tau_m=3$ and $70$ yr).
A shorter time scale ($\tau_m=3$ yr) predicts a distribution that is more shifted towards negative values, indicating faster turnover and hence larger decay, while a larger $\tau_m=70$ yr predicts smaller decay than the data (Fig. \ref{fig:gbm_marginals}c). The inferred model can correctly predict the full distribution of clonal counts at a timepoint $t_2$, given its abundance at an earlier timepoint $t_1$ (Fig. \ref{fig:gbm_marginals}d-e).
The model generalizes well to held-out samples (i.e. not used for inference) of the same individual: the distribution of counts on a third time point conditioned on counts on the first two shows excellent agreement with the data (Fig. \ref{fig:gbm_marginals}f-g).
\rev{The model inferred on donor 2 of {\em Dataset A} is also successful in predicting the marginals of donor 1 (\SMRef{Fig. S8a}), consistent with the fact that their inferred timescales are very similar (Table \ref{tab:params_gbm}).
This gives us another test of robustness and suggests that the inferred dynamical models generalize across people.
}

\begin{figure}
	\centering
	\includegraphics[width=1\linewidth]{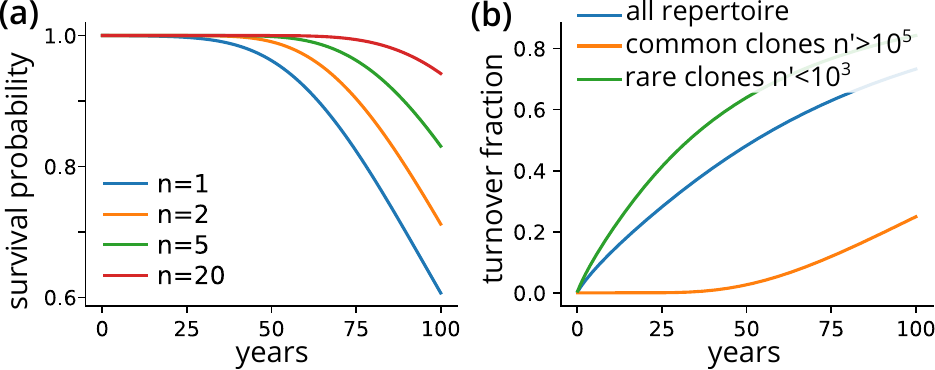}
	\caption{
		{\bf Probability of future survival/turnover.}
		(a): Survival probability of a clone after $t$ years given an experimental observation $n$, \SMRef{Eq. S11}. We have used $\tau = 14.5$, $\theta=9$, $N_{cell}=10^{10}$, $n_0=10^4$, and $M=10^6$.
		(b): Fraction of clones that \rev{go extinct} after
                $t$ years (\SMRef{Eq. S10} of Supplementary Material), computed with the same parameters as before.
		The three lines consider clones in different ranges of abundances as specified by the legend.
	}
	\label{fig:gbm_survival}
\end{figure}

A characteristic decay time of $\tau_m\approx 15$ years implies that clones can, in fact, survive for much longer than that. To see this, we can use the model to compute the probability of survival of a clone (i.e. that there remains at least one cell from that clone in the organism) over time, as a function of its cell count in a given repertoire sequencing experiment.
Strikingly, after 25 years, even clones represented by just a single cell in the experiment ($n=1$) are predicted to survive almost certainly, while clones with larger counts will very likely never go extinct in our lifetime (Fig. \ref{fig:gbm_survival}~a). The reason for this long survival, despite a more modest decay time, is that experiments only capture a very small fraction of all memory B cells ($10^5$ out of $\approx \rev{10^{10} \textendash 10^{11}}$). Even clones with a single cell in the sampled repertoire are typically abundant in the whole repertoire \rev{(average count $\approx 10^4 \textendash 10^5$, using the parameters of Fig. \ref{fig:gbm_survival})}. Even with decay, they are likely to survive.
The framework also allows us to compute the probability of observing a clone in future experiments as a function of its count in the current one (\SMRef{Fig. \rev{S8b}}).

We can use the model to answer a related question: after $t$ years, what is the fraction of clones that will go extinct and be replaced by newly generated ones?
Fig. \ref{fig:gbm_survival}b shows that repertoire renewal is quite slow, with replacement of half the clones after $50$ years (blue line for the whole repertoire).
The turnover depends on clone abundances: large clones (with total number of cells in the organism $> 10^6$, which are likely sampled in an experiment, show negligible turnover before 50 years.
Even taking the lowest estimate of $\tau_m = 5$ yr from {\em Dataset B}, half of the repertoire would be renewed after around $25$ years (\SMRef{Fig. \rev{S8c}}). Overall, these estimates suggest that the memory B cell repertoire is stable over decades.

\subsection{Rate and size of new memory B cell clones, and their total diversity}
\label{sec:memory_n0}

In addition to $\tau_m$ and $\theta_m$, we inferred two additional parameters of the Geometric Brownian Motion: $s$, the rate of introduction of a new memory B cell clone, and $n_0$, the initial number of cells in each new clone. The rate $s$ provides an upper bound on the rate with which new germinal centers (GC) appear, since each GC may produce multiple clones. However, if we assume that all cells from a GC belong to the same clone \cite{victora2022germinal}, then $s$ may be directly interpreted as the rate of appearance of new productive GCs.

We found that the absolute initial memory clone size $n_0$ was underdetermined under our statistical approach. However, the ratio of initial clone size to the total number of memory cells, $n_0/N_{\rm cell}$, is well inferred (\SMRef{Fig. S6b}). We can thus infer $n_0$ by using estimates of the total number of memory B cells \rev{$N_{\rm cell} \approx 10^{10} \textendash 10^{11}$} \cite{lythe2016many, seifert2016human, sender2023total}, for which the GBM model makes a mathematical prediction as a function of the 4 model parameters ($\tau_m$, $\theta_m$, $s$, $n_0$), see \SMRef{Sec. S3}. That additional constraint allows us to obtain $n_0$ as a function of the other 3 parameters, which we inferred reliably. The results of this procedure, reported in Table \ref{tab:params_gbm},
lead to an estimated $n_0 \sim 10^4$ memory cells per new clone produced by a germinal center.
\rev{Recall that $n_0$ is an average, as the output of germinal centers is known to be highly variable \cite{matz2023persistent}.} 
Note that \rev{$n_0$} implies an initial frequency of new clones of less than one part in a million (\rev{$10^4/10^{10}=10^{-6}$}), meaning that new clones are typically not expected to be found by a repertoire sequencing experiment. This explains why $n_0$ is hard to infer from the data.

The rate of appearance of new memory clones, $s$, is estimated to be $10^3 \textendash 10^4$ per year. Assuming that each GC produces only one or a few clones, that implies thousands of GC formed every year.
These estimates are obtained for {\em Dataset A} only, as {\em Dataset B} only allows us to put an upper bound on it.

By deriving exact analytical formulas from the model (see \SMRef{SI Sec. S3A}), we can estimate the real diversity $N_{\rm clone}$ of clones, i.e. the total number of memory B cell clones in the organism. It can be expressed as a mathematical function of the 4 parameters, yielding an estimate of $N_{\rm clone}\approx 10^6$.

Interestingly, our data on 5 individuals suggests that the ratio of initial clone size to the total number of memory cells $n_0/N_{\rm cell}$ increases with age, and that the total diversity $N_{\rm clone}$ decreases as a result (\SMRef{Fig. S9}).
This would mean that younger individuals either have a larger pool of memory cells (higher $N_{\rm cell}$), or that they produce fewer memory cells from each germinal center, resulting in higher diversity.
This last observation would be in agreement with previous work on age-dependent diversity of B-cells \cite{tabibian2016aging}.
However, these conclusions are based on only 5 points, and there can be dataset-dependent biases, requiring confirmation from future analysis of other datasets.

\subsection{1\% of memory B cells differentiate into plasmablasts every year}
\label{sec:plasmablast}

\begin{table}
	\begin{tabular}{c||c|c|c} 
		& differentiation & decay time & \rev{fluctuation} \\
		& rate $\rho$ (1/year) & $\tau_p$ (year) & \rev{time} $\theta_p$ (year) \\ [0.5ex] 
		\hline
		& & & \\[-1.8ex]
		donor 1 (A) & $0.003\textendash0.03$ & $0.5\textendash3$ & $0.1\textendash0.3$ \\
		donor 2 (A) & $0.01\textendash0.1$ & $0.25\textendash1$ & $0.05\textendash0.1$ \\ [1ex] 
	\end{tabular}
	\caption{
		\textbf{Inferred parameters of the plasmablast-memory coupled dynamics.}
		The parameters are learnable only for {\em Dataset
                 \rev{A}}.
		The inference uses the memory parameters learned at the previous step, in Tab. \ref{tab:params_gbm}.
		Each parameter shows a range that depends on the unknown variability of the ratio of plasmablasts to memory cells, $r \sim 0.01 \textendash 0.1$, and the fraction of plasmablasts that differentiate from memory cells with respect to the total plasmablasts, $f \sim 0.3 \textendash 0.8$.
	}.
	\label{tab:params_pb}
\end{table}

We next applied our inference approach to the plasmablast repertoires. We only managed to successfully perform that inference for {\em Dataset A}, likely because stronger noise in {\em Dataset B} impeded learning (see \SMRef{Sec. S4} and \SMRef{Fig. S10} for details).
The model now has three learnable parameters: the decay time $\tau_p$, the \rev{fluctuation time} $\theta_p$, and the rate of imports from the memory pool $\rho$ (see above and Methods).
Because repertoire sequencing only allows us to capture relative frequencies and not absolute numbers, the inference scheme leaves two unidentifiable quantities by construction (see Methods):
the fraction of plasmablasts that originate from memory cells, $\phi$, and the ratio between the number of plasmablasts and memory cells, $r$.
We observe that around 25-30\% of sampled plasmablasts can be assigned to clones that are represented in the sampled memory pool, setting a lower bound $\phi\geq 30\%$.
Because the sampling of the memory pool is very limited, the real value of $\phi$ is likely much larger. To account for this uncertainty, we performed the inference fixing $\phi$ to values ranging from $30\%$ to $80\%$.
The ratio $r$ of plasmablast to memory cells depends on many factors, and probably varies in time, but rough estimates from the literature \cite{carsetti2022comprehensive} as well as the ratio of plasmablast to memory in experiments suggest a value ranging from $r=1\%$ to $r=10\%$.
Results reported in Table \ref{tab:params_pb} show a range that is a consequence of this variability and the error associated with the maximum likelihood procedure.
The inferred values of the 3 parameters $(\rho,\tau_p,\theta_p)$ while fixing $\phi$ and $r$ to various values are reported in \SMRef{Fig. S11}.

The memory-to-plasmablast differentiation rate is found to be around $0.01$/year or larger: on average, at least $1$ out of $100$ memory cells differentiates into a plasmablast each year. 
The net decay time of plasmablast clones ranges from a few months to a few years, which is 10 to 100 times faster than the $15$ years of the memory B cell clones.
Note that this decay time is still large compared with the life span of plasmablasts, which is a few days \cite{khodadadi2019maintenance}. This implies that clone sizes must be maintained by cell proliferation to counter cell death, consistent with reported proliferation activity in plasmablasts \cite{nutt2015generation}.
The fluctuation time $\theta_p$ is small compared to the decay time $\tau_p$, meaning that fluctuations in that proliferation activity are large. These large variations may originate from clonal heterogeneity in the cell differentiation, division, and death rates, which is not explicitly described by our model but would effectively lead to large fluctuations.

\begin{figure}
	\centering
	\includegraphics[width=\linewidth]{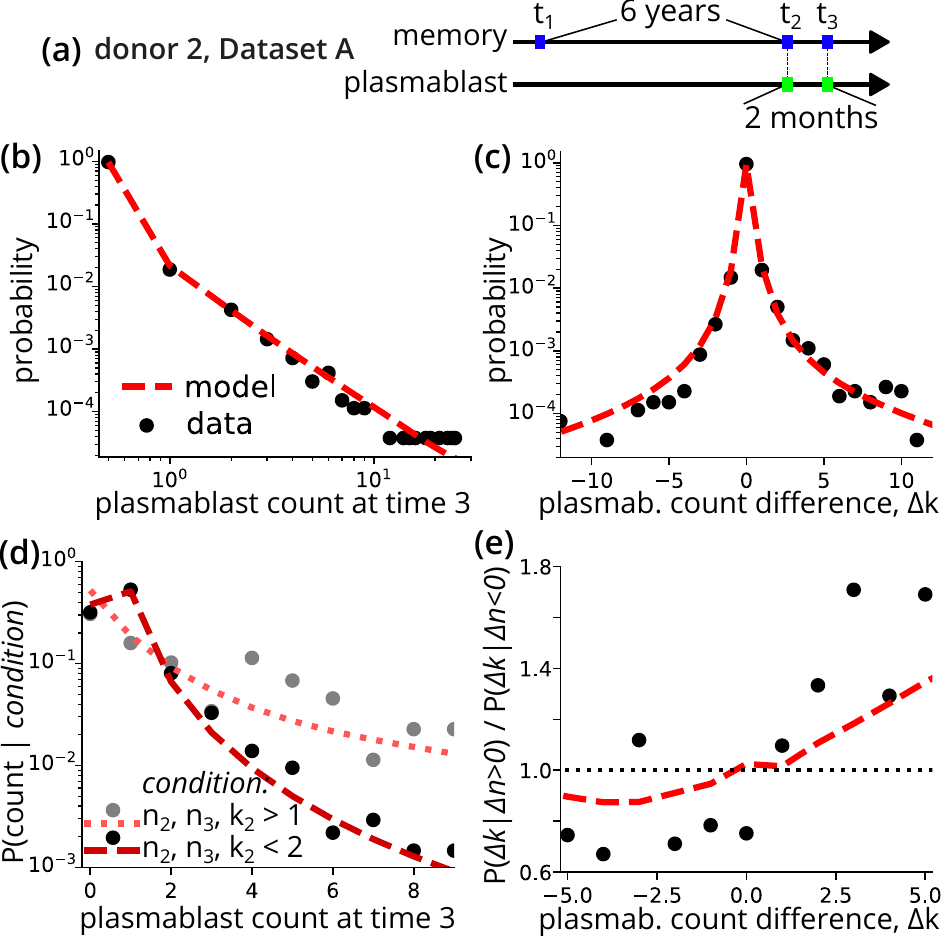}
	\caption{
          {\bf Validation of the plasmablast-memory coupled dynamics.}
		Conditional distributions of plasmablast counts from donor 2 from {\em Dataset A}.
		The model uses the best-fit parameters inferred for $r=0.02$ and $\phi=0.5$, which are $\tau_p=0.4$ and $\theta_p=0.06$ years, implying a differentiation rate $\rho=0.026$/year.
        (a) Summary of the samples and timepoints of donor 2.
		(b) Plasmablast clone size distribution at the third time point, conditioned on the clone being present in the memory sample at the first time point.
		(c) Distribution of the count difference $\Delta k$ between $t_2$ and $t_3$.
		(d) Distribution of plasmablast counts at $t_3$ given a condition on the second and third memory-sample counts (labeled by $n$) and on the second plasmablast-sample counts (labeled by $k$). \rev{Black and gray points refer to the two different conditions.}
		\rev{For the second condition (black points), we select only plasmablasts with clones also present in at least one of the memory samples, i.e., we select only the plasmablasts that are memory-derived (this is automatically satisfied for the first condition).}
		(e) Odds ratio of two probability distributions over the plasmablast count difference $\Delta k$. The numerator is conditioned on an increase in the corresponding memory counts, $\Delta n = n_3 - n_2 > 0$, and the second to decrease $\Delta n < 0$.
		A ratio larger than 1 means that clones with a given $\Delta k$ are more likely to be observed when the memory counts grow.
	}
	\label{fig:pb_marginals}
      \end{figure}

We validated the model by comparing its predictions to data, focusing on the model inferred from {\em Dataset A}'s second donor (Fig. \ref{fig:pb_marginals}a). The plasmablast clone size distribution (Fig. \ref{fig:pb_marginals}b) is well reproduced, as well as the distribution of fold changes over 2 months (Fig. \ref{fig:pb_marginals}c). Because both $\tau_p$ and $\theta_p$ are short relative to the memory parameters, the turnover dynamics of plasmablasts is fast, meaning that the plasmablast repertoire is a slightly delayed mirror image of the memory one. Hence, the power law observed in its clone size distribution is largely inherited from that of the memory pool.

In addition, the model correctly predicts the distribution of plasmablast clone sizes at the third time point given their recorded abundance in other samples (Fig. \ref{fig:pb_marginals}d), including the memory pool.
Lastly, the model makes a specific prediction about how the change of clone size in the plasmablast pool is influenced by the change of clone size in the memory pool of the same clone, which agrees with the data (Fig. \ref{fig:pb_marginals}e): clones that grew in the memory pool are more likely to grow in the plasmablast pool in the same interval ($p = 3.6 \; 10^{-9}$, Mann-Whitney U test).

\subsection{\rev{Robustness of the inference}}
\label{sec:robust}

\rev{
	A limitation of our approach is its dependence on the choice of software for reconstructing clonal families.
	To check for robustness to that choice, we repeated the analysis using the alternative ChangeO tool instead of HILARy (see Methods) for donor 2 of \emph{Dataset A}.
	 Inferring the noise model on these families yielded similar parameters (Fig. \SMRef{S12}a). The likelihood of the Poisson and the linear Negative Binomial models was almost the same (Fig. \SMRef{S12}b), justifying the use of the Poisson model for downstream analysis. Similarly, the parameters of the memory and plasmablast models inferred with the ChangeO families showed negligible differences with HILARy (\SMRef{Table S1}). This indicates that our inferred parameters are insensitive to the choice of clustering algorithm.

	In general, clustering algorithms can wrongly merge together distinct clones and thus slightly overestimate their decay time $\tau_m$. To control for this effect, we repeated the analysis by defining a \quotes{clone} as a set of identical heavy-chain sequences. Under that definition, any hypermutation creates a new clone. The inferred decay time $\tau_m$ can thus be seen as a lower bound. Applying inference with that definition indeed yields a smaller $\tau_m$ than with the HILARy clones, but still large ($\sim 10$ years, see \SMRef{Table S2}), confirming our main conclusions.
	
	Finally, the analysis relies on the FACS sorting that can wrongly sort some B cells between memory cells and plasmablasts. To assess the impact of mis-sorting, we artificially replaced a fraction of memory cells by plasmablasts collected at the same time point and repeated the inference of the GBM (\SMRef{Fig. S13}).
	As expected, the inferred average decay time $\tau_m$ decreases as the fraction of plasmablasts in the sample increases, since plasmablast dynamics are faster. However, this change is only visible when a large fraction (more than $10\%$) of cells are mis-sorted, making our conclusions also robust to that potential source of errors.	
}

\section{Discussion}
\label{sec:discussion}

We presented a framework to quantify the dynamical changes of memory B-cell and memory-derived plasmablast repertoire in healthy donors. Our approach infers the parameters of stochastic dynamics models while accounting for experimental noise.
The results give us access to biologically informative quantities and provide an accurate statistical description of clone counts in longitudinal repertoire sequencing experiments.

The method is, however, reliant on the quality and structure of the dataset. It requires experimental replicates to infer a noise model, which is particular to each protocol and laboratory setting. We found that two distinct models, Poisson and Negative Binomial, best explained the 2 datasets we analyzed.
Another requisite is that longitudinal samples be properly spaced.
For example, the one-year time window of {\em Dataset B} was not enough to learn the longer degradation timescale of memory cells.
Also, experimental noise and sequencing depth are important: the large negative binomial noise of {\em Dataset B} concealed the signal of the plasmablast dynamics, preventing inference in that case.\\

By modeling the memory cell dynamics with a Geometric Brownian Motion, we accurately captured the statistics of clone counts in samples, including in future ones not used for inference.
The inferred parameters teach us that the half-time of memory cell clones is around 10 years.
This implies that clones that are observed in experiments survive for tens of years. 
{This number differs from existing measurements on the lifetime and turnover of lymphocytes \cite{DeBoer2003a,cosgrove2021hematopoiesis, krueger2017t, sender2021distribution}, which instead describe the life expectancy of {\em single cells}.
This long lifespan of clones arguably correlates with immune memory and is consistent with the observed time scale of protection against encountered pathogens \cite{amanna2007duration, crotty2003cutting}.
A previous analysis of T-cell clonal lifespans led to estimates of the same order of magnitude, from several years to decades \cite{bensouda2023inferring}, and increased with age, consistent with the reduction of thymic output. Because of the limited number of individuals, our analysis did not allow us to probe the age dependence of memory B cell clonal survival.}
Because the half-life of 10 years is much larger than the lifespan of a single memory cell, which is only a few months  \cite{westera2015lymphocyte, sender2021distribution}, our estimates imply that the maintenance of memory clones must involve cell proliferation to compensate for cell death, rather than through the existence of long-lived quiescent cells \cite{cancro2021memory, lam2024guide}. In that picture, individual cells divide and die quickly, with the two effects balancing each other and keeping the clone size stable for years.
As a result, repertoire turnover, i.e. the replacement of old memory clones by new ones, is relatively slow and mostly affects small clones that are not sampled in experiments.

The inference also gives us access to parameters connected with the emergence of memory clones and, therefore, with properties of germinal centers.
We inferred that every year around 1,000-10,000 new memory clones appear, each one of them producing $10,000$ memory B cells, and that there are around $N \sim 10^6$ different memory clones in our body. 
It is difficult to compare these numbers to direct estimates from the literature. However, we can use the parameters used in standard models of affinity maturation \cite{molari2020quantitative} to get a rough estimate of germinal center output.
Starting from a few thousand cells in a germinal center, after two rounds of replication twice a day, a few tens of thousands of cells are tested against the antigen each day.
Of those, between 10\% and 33\% pass the selection step, which then have around a 5\% probability of differentiating into memory cells, making a few hundred new cells per day.
After a few weeks, the memory cell output would be around $10^4$, consistent with our estimate.
These GC-related numbers, however, should be taken with a grain of salt: we expect them to mostly affect the statistics of small clones, which by definition are not well sampled by the repertoire sequencing experiments, making their inference potentially unreliable. Our estimates also rely on the assumption that the dynamics of germinal centers are fast compared to the \rev{timescales} considered for repertoire evolution. However, germinal \rev{centers} can last for a few weeks and in some cases for much longer periods \cite{mesin2016germinal}, and can be seeded by existing memory clones.

Our approach applied to memory-derived plasmablasts infers \rev{that, over the course of a year, $1\%$ of memory cells differentiate into plasmablasts.}
This relatively high rate suggests a scenario in which differentiation is not just the result of isolated events triggered by antigen restimulation but is instead continual and widespread \cite{bernasconi2002maintenance, horns2020memory, phad2022clonal}.
However, the process is clone- and thus probably antigen-dependent, as evidenced by the large fluctuations that we inferred (small $\theta_p$).
The model, however, ignores possible variability and heterogeneity in the differentiation rate across clones and time, \rev{as well as heterogeneity in clone degradation of both memory cells and plasmablasts, and folds these variations into the parameters $\theta_m$ and $\theta_p$. More realistically, the fate of clones and their ability to proliferate and differentiate into plasmablasts depends on their initial affinity to the antigen, as well as the full history of their affinity maturation. Large, long-lived clones may be driven by a variety of slightly different antigens, e.g. variants of the same pathogen that evolve over years or even decades, like respiratory diseases such as influenza or SARS-CoV-2. More refined, antigen-aware analyses would require reliable labeling of antibody lineages by antigen specificity, which is currently impractical, both experimentally and computationally. The model does not account for the possible coupling between the rates of proliferation and differentiation that one could expect from antigen-dependent stimulation.
However, such an effect would likely be obscured by correlations between proliferation and differentiation that are already implied by our model over long timescales: clones that have proliferated a lot tend to be bigger and to have produced more plasmablasts, even at a constant differentiation rate. Additional coupling between proliferation and differentiation arguably occurs at smaller timescales.
Future models could include and infer such effects using data with shorter sampling intervals.
}

{Phad et al~\cite{phad2022clonal} report the stability of memory B cell clonal families over several years. The argument is based on the observation that the clonal overlap between two samples from the same person taken at different times is not much reduced compared to that between two samples taken at the same time. Mikelov et al~\cite{mikelov2022memory} also report the stability of memory B cell clones on the year timescale of their experiments based on repertoire overlap. These arguments are indirect and limited to the timescale of the experiment.  Here we show that this stability extends for much longer than several years, identifying a characteristic decay time of 15 years. This implies long lifespans of several decades or more for sampled clones, much longer than the span of the original experiments. We also show that this timescale is the same for both individuals, even though they were sampled 6 and 10 years apart, and is consistent with our analysis of data from Mikelov~\cite{mikelov2022memory}, whose donors were sampled 1 year apart, suggesting universality of this timescale. While previous work~\cite{phad2022clonal} reports a significant decrease of the number of shared clones for small compared to large clones in samples collected 6 and 10 years apart, we show that this decrease is largely due to sampling biases. We predict that even sampled singletons likely belong to clones large enough to persist over $\sim 20$ years. In summary, our analysis allows us to put concrete values on the lifetime of clones and memory-to-plasmablast differentiation, extending the list of precise numbers known in biology and immunology~\cite{milo2010bionumbers,cosgrove2021hematopoiesis}, characterizing processes that occur on timescales longer than sampled by experiments. The model also allows us to estimate other biologically relevant quantities that are obtainable only through an inference-based approach, such as repertoire turnover,  cell proliferation, and germinal-center-related events.
Phad et al~\cite{phad2022clonal} estimate that  $20\%$ of circulating plasmablasts derive from long term memory B cell families, based on the clonal overlap between the two subsets. Our inference approach revealed that this number ($\phi$ in our notation), cannot be estimated accurately from data. However, we can put a lower bound of $30\%$.
They speculate that sustained plasmablast production may be antigen independent and come from polyclonaly stimulated memory B cells. Our analysis also shows sustained and continuous plasmablast production, however the large heterogeneity in future clone sizes suggests external differentiating factors.
}

A possible extension of our framework can include sub-populations and could be interesting in the context of memory-cell stability. For example, in \cite{mikelov2022memory}, from which {\em Dataset B} is derived, two clusters of low and high temporal persistence were identified.
Our analysis ignored these two subgroups, characterizing only the average behavior of the repertoire.
Generalizing our approach to multiple subpopulations would increase the number of parameters and the complexity of the model, making the inference harder and leaving some parameters potentially unidentifiable.
One could also determine subgroups based on objective observables, such as the isotype, gene expression signatures, sequence features of the B-cell receptor, or phylogenetic properties of the clonal family, run the inference on each subgroup, and compare the learned parameters across them. {The method could also be applied to antigen-specific subsets, such as measles, tetanus and influenza clones reported in Phad et al~\cite{phad2022clonal}, given that these subsets are large enough, and used to predict the evolution of the corresponding antibody titers.}

The models inferred using our method reproduce very accurately the statistics of longitudinal repertoire data in healthy conditions. This suggests the possibility of using them as null models to detect and describe pathological or unexpected variations.
This approach could be integrated into methods to study clonal expansion and subsequent contraction following immune stimuli, including vaccination \cite{galson2016b, horns2020memory, hoehn2021cutting}, or to characterize the unique dynamics of clones interacting with chronic diseases such as HIV \cite{nourmohammad2019fierce, mazzolini2022inspecting}.

\section{Methods}
\label{sec:methods}

\subsection{\rev{Inference of clonal families}}

\rev{
To group sequences originating from the same naive ancestor into clones, we mainly used the HILARy software \cite{spisak2024combining}. 
Its computation uses both the heavy and light chains when the pairing is accessible (\emph{Dataset A}), while only the heavy chains for \emph{Dataset B}.
Compared to standard software, HILARy uses information on mutations in the V-gene region and optimizes the clustering threshold for each class defined by V,J gene usage and a CDR3 length.
Since we need to track longitudinally the same clone at different time points and for different cell types (memory or plasmablast), we performed clone clustering by merging all the samples of the same donor.

To test the robustness of the inference with respect to changes of the clonal assignment algorithm, we also repeated part of the analysis using the software provided by the Change-O toolkit \cite{gupta2015change}, using a threshold of 0.15 on the length-normalized Hamming distance.
}

\subsection{Inferring a dynamical model from noisy experimental samples}

We denote the list of true clonal abundances at a given time point $t_k$ by the vector $\mathbf{n}'^k = \{ {n'}^k_1, {n'}^k_2, \ldots \}$.
The corresponding frequencies read $f'_i = n'_i / N_{\rm cell}$, where $N_{\rm cell} = \sum_i n'_i$ is the total number of cells in the body.
The temporal evolution of these abundances is specified by a stochastic dynamical model that depends on some parameters ${\bf p}_{\rm dyn}$.
The model gives, for the clone $i$, the probability of observing its frequencies at $t_1$ and $t_2$: $P_{\rm dyn}({f'}_i^1, {f'}_i^2 | {\bf p}_{\rm dyn})$.
Since we are interested in studying long term dynamics and not a response to acute stimulations, we assume that the model is in a stationary state and that clones are independent.

Our knowledge about count variations comes from longitudinal sequencing experiments, which provide a very small fraction of the immune repertoire.
We define a noise model that provides the experimental count $n_i$ given a real frequency: $P_{\rm noise}(n_i | f'_i, {\bf p}_{\rm noise})$.
In these notations and throughout, the prime indicates true values for the entire organism, while experimental observations are labeled without prime.

Based on the approach of previous work \cite{puelma2020inferring, bensouda2023inferring, koraichi2022noiset}, the inference scheme combines the dynamical model and the noise model to write down the probability of observing experimental counts at different time points by integrating over the true frequencies $f'$:
\begin{equation}
	\begin{aligned}
	P(n^1_i, n^2_i | {\bf p}_{\rm noise}, {\bf p}_{\rm dyn}) = \int d {f'}^1 d {f'}^2 P_{\rm dyn}({f'}^1, {f'}^2 | {\bf p}_{\rm dyn}) 
	\\
	 \times P_{\rm noise}(n^1_i | {f'}^1, {\bf p}_{\rm noise}) P_{\rm noise}(n^2_i | {f'}^2, {\bf p}_{\rm noise}) ,
	\end{aligned}
	\label{eq:prob_exp_counts}
\end{equation}
Using this probability, under the assumption of independent clones, one can then write the log-likelihood of observing the dataset at the different time points:
\begin{equation}
	\mathcal{L}( {\bf p}_{\rm noise}, {\bf p}_{\rm dyn}) = \sum_i \log P(n^1_i, n^2_i | n^1_i + n^2_i > 0,  {\bf p}_{\rm noise}, {\bf p}_{\rm dyn}) ,
	\label{eq:ll}
\end{equation}
where the probabilities are conditioned on clone $i$ to be present in the dataset.
The maximization of the likelihood allows us to find the best set of parameters that describe the data.
Practically, we first learn the noise parameters ${\bf p}_{\rm noise}$ by using experimental replicates as explained in the next section, and, secondly, we learn the parameters of the dynamics ${\bf p}_{\rm dyn}$ by plugging the optimal ${\bf p}_{\rm noise}^*$ into Eq. \ref{eq:prob_exp_counts}.

\begin{table}
	\centering
	\begin{tabular}{c|c}
		$N_{\rm cell}$ & Number of memory B-cells in the body \\ \hline
		$M$ & Number of cells in the experiment \\ \hline
		$N_{\rm clone}$ & Number of memory clones in the body \\ \hline
		$\alpha$ & Power law exponent of the cumulative clone count distribution \\ \hline
		$f_{\rm min}$ & Minimal clonal frequency $= 1/N_{\rm cell}$ \\ \hline
		$a$, $b$ & Negative Binomial noise parameters (Eq. \ref{eq:var_negbin}) \\ \hline
		$\tau_m$ & Decay time of the GBM for memory clones \\ \hline
		$\theta_m$ & Fluctuation time of the GBM for memory clones \\ \hline
		$s$ & Rate of introduction of new memory clones \\ \hline
		$n_0$ & Initial number of cells of a new clone \\ \hline
		$\rho$ & Differentiation rate from memory to plasmablast \\ \hline
		$\tau_p$ & Decay time of the GBM for plasmablasts \\ \hline
		$\theta_p$ & Fluctuation time of the GBM for plasmablasts \\ \hline
		$\phi$ & Fraction of plasmablasts with memory origin \\ \hline
		$r$ & Plasmablast to memory cell ratio in the body \\
	\end{tabular}
	\caption{\rev{List of parameters.}}
	\label{tab:math_params}
\end{table}

\subsection{Noise models}
\label{sec:methods_noise}

Because repertoire sequencing experiments are based on samples representing a small part of the entire organism, and because of experimental noise, experimental counts are a noisy function of the true clonal frequencies.
We describe this noisy relationship through three alternative models that we present in order of complexity, i.e., the number of adjustable parameters they have.
In a pure sampling regime, the experimental-count probability of the clone $i$ is given by a Poisson distribution with average $\mu_i = M f_i'$, where $M = \sum_i n_i$ is the total number of counts in the experiment (single cell, reads, or unique molecular identifier, depending on the case).
This parameter-free model describes a scenario in which there is no other noise than from sampling.
The other two models are based on the negative binomial distribution, commonly used in the presence of amplification biases in RNA sequencing data \cite{love2014moderated} and repertoire sequencing \cite{puelma2020inferring, koraichi2022noiset}.
We parametrize this distribution by fixing the mean to $\mu_i = M f_i'$ as before and writing down the variance $\sigma^2_i$ as an increasing function of the mean, with two free parameters $a$ and $b$: 
\begin{equation}
	{\sigma_i^2} = \mu_i + a (\mu_i)^b .
	\label{eq:var_negbin}
\end{equation}
When $a= 0$ we get back the Poisson distribution.
We consider two options: one in which the exponent $b$ is fixed to $1$, i.e., the variance is proportional to the average with a factor of $1 + a$, and the full model with the two parameters $(a,b)$ to learn.

This noise model is inferred by using experimental replicates performed at the same time point.
Given the experimental count $n_i^r$ for the clone $i$ in the replicate $r$, we can write down the probability of observing the clone in two different replicates:
\begin{equation}
	\begin{aligned}
		P(n^1_i, n^2_i | {\bf p}_{\rm stat} ,{\bf p}_{\rm noise}) = \int d f' \; P_{\rm stat}(f' | {\bf p}_{\rm stat})
		\\
		\times P_{\rm noise}(n^1_i | \mu=f'M, {\bf p}_{\rm noise}) P_{\rm noise}(n^2_i | \mu=f'M, {\bf p}_{\rm noise}) ,
	\end{aligned}
	\label{eq:prob_exp_counts_repl}
\end{equation}
where $P_{\rm stat}(n | {\bf p}_{\rm stat})$ is the stationary distribution for the repertoire abundances, which
we assume to be power-law distributed, $P_{\rm stat}(f' | {\bf p}_{\rm stat}) \propto 1/f'^{\alpha+1}$, in agreement with experimental observations \cite{jiang2013lineage, weinstein2009high} and previous analyses \cite{desponds2016fluctuating, chardes2022affinity, bensouda2023inferring}.
This distribution depends on two free parameters: the power law exponent $\alpha$ and a minimal frequency $f_{\rm min}$ to ensure integrability.
By using the Eq. \ref{eq:prob_exp_counts_repl}, one can obtain the likelihood of observing a set of replicates using Eq. \ref{eq:prob_exp_counts} and Eq. \ref{eq:ll} but with $t_1=t_2$.
\SMRef{Sec S2} provides specific details for its computation and maximization.

\subsection{Geometric Brownian Motion for memory \rev{cell} dynamics}
\label{sec:methods_GBM}

The minimal model to describe the dynamics of memory B cells in healthy condition is the Geometric Brownian Motion (GBM), according to which abundances of cells decrease exponentially on average and are subject to geometric fluctuations:
\begin{equation}
	\frac{d n'_i}{dt}  = - \frac{n'_i}{\tau_m} + \frac{n'_i}{\sqrt{\theta_m}} \eta_i(t) ,
	\label{eq:GBM}
\end{equation}
where $\tau_m$ is the \rev{timescale} for the average exponential degradation, $\theta_m$ quantifies the amplitude of fluctuations, and $\eta_i(t)$ is a normalized Gaussian white noise implemented with the It\^o convention.
Clones go extinct when they hit the absorbing boundary $n' = 1$, and new clones are introduced with a constant rate, $s$, starting with an initial abundance $n_0$.
The balance of extinction and new introduction leads to a stationary condition, with an abundance distribution that has a power law tail $p(n') \propto 1/n'^{\alpha + 1}$ (\SMRef{see \cite{desponds2016fluctuating} and Sec. S2E}), where the exponent is connected with the model parameters through $\alpha = 2 \theta_m / \tau_m$.
The power law behavior and its simplicity are the main reasons for choosing this model. 
In addition, it has been shown that when observing two time points sufficiently far apart in time, i.e. longer than a few weeks, more refined models, including antigen recognition and clonal expansion, can be well approximated by the GBM \cite{desponds2016fluctuating}.

Eq. \ref{eq:GBM} defines a dynamical model $P_{\rm dyn}$ (\SMRef{see Sec. S2E}).
Combining it with the noise model $P_{\rm noise}$ allows us to run the inference machinery described above (Eqs. \ref{eq:prob_exp_counts},  \ref{eq:ll}) and to infer the biologically informative parameters of the GBM, as discussed in Sec. \ref{sec:memory_tau}.

\subsection{Coupling \rev{plasmablast} and memory \rev{cell} dynamics}
\label{sec:methods_smem_pb}

Memory cells can differentiate into plasmablasts, providing a rapid response against pathogens encountered in previous infections.
We denote by $\rho$ this differentiation rate, and we let the dynamics of the number of plasmablasts in the clone $i$, $k'_i$, follow a Geometric Brownian Motion as for memory cells:
\begin{equation}
	\frac{d k'_i}{dt}  = \rho n'_i - \frac{k'_i}{\tau_p} + \frac{k'_i}{\sqrt{\theta_p}} \xi_i(t) .
	\label{eq:mem_pb}
\end{equation}
Plasmablast clones degrade exponentially on average with a time $\tau_p$ and are subject to fold fluctuations over time scale $\theta_p$.
By coupling Eqs. \ref{eq:GBM} and \ref{eq:mem_pb}, one obtains a dynamical model and can learn the plasmablast parameters given the memory parameters learned previously.
It turns out that it is not possible to learn the three new parameters simultaneously. This can be seen by writing the equation for the true frequencies $g'_i = k'_i / K'_m$, where $K'_m$ is the total number of plasmablasts generated by memory cells, and observing that it loses the dependency on $\rho$ (see \SMRef{Sec. S4}).
This indeterminacy can be solved by finding the relation between the total number of cells at stationary state, $K'_m = \rho \tau_p N_{\rm cell}$, and comparing it with the empirical ratio of these two numbers.
In particular, we call $\phi = K'_m / K'$ the fraction of plasmablasts that differentiated from memory cells compared to the total population of plasmablasts $K'$ (some plasmablasts can differentiate from naive cells), and $r = K'/N_{\rm cell}$ the plasmablast-to-memory ratio.
This leads to $K'_m = r \phi N_{\rm cell}$, and $\rho = r \phi / \tau_p$, where $\tau_p$ is one of the parameters to fit in the optimization of the likelihood, and $r$ and $\phi$ are directly given by empirical estimates from the repertoires.

\subsection{Code Availability}

The code used for processing the immune-sequencing repertoire datasets, for performing the inference, and for generating the figures is available at the repository \url{https://github.com/statbiophys/memory_plasmab_dynamics}

\section*{Acknowledgements}
This work was partially supported by the ANR-
19-CE45-0018 “RESP-REP" and ANR-22-CE95-0005-01 ``DISTANT'' grants  from the Agence Nationale
de la Recherche.

\bibliographystyle{pnas-new}

\onecolumngrid

\newpage\hbox{}\thispagestyle{empty}

\begin{center}
	\textbf{\large SUPPLEMENTARY MATERIAL}
\end{center}

\setcounter{figure}{0}
\setcounter{section}{0}
\setcounter{table}{0}
\setcounter{equation}{0}

\renewcommand{\figurename}{Supplementary Figure}
\renewcommand{\thesection}{Supplementary Note \arabic{section}}
\renewcommand{\thefigure}{S\arabic{figure}}
\renewcommand{\thesection}{S\arabic{section}}
\renewcommand{\tablename}{Supplementary Table}
\renewcommand{\thetable}{S\arabic{table}}
\renewcommand{\theequation}{S\arabic{equation}}

\section{Datasets}

\subsection{Single-cell {\em Dataset A}}
\label{sec:SM_data}

\begin{figure}[htbp]
	\centering
	\includegraphics[width=1\linewidth]{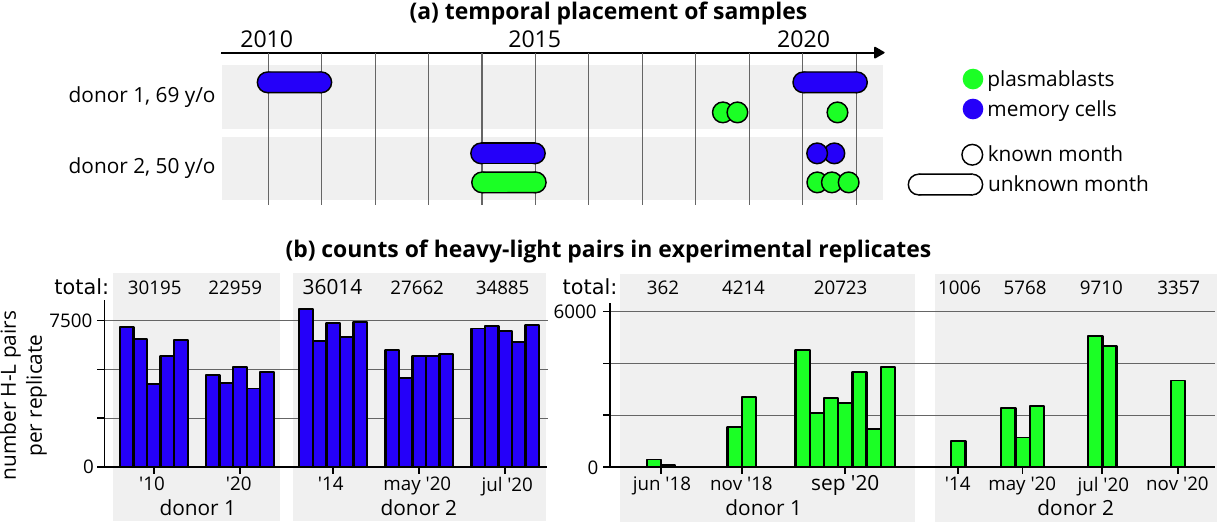}
	\caption{{\bf \em Dataset A}  \cite{phad2022clonal}
		(a): dataset samples of the two donors for memory cells (blue) and plasmablasts (green). 		For some samples the metadata provides the year and month of the experiment (circles), while for others only the year (elongated points). 
		(b): counts of cells that contain a heavy-light sequence for each experimental replicates of each sample. 
		On the left memory cell samples (blue), on the right plasmablasts (green).
	}
	\label{fig:data1}
\end{figure}

The single-cell immune sequencing {\em Dataset A} is described in \cite{phad2022clonal}. B cells were obtained with the high-throughput
single-cell 10X Genomics platform and available under ArrayExpress
accession \textit{E-MTAB-11174} and \textit{E-MTAB-11697}.
This dataset consists of longitudinal sequences of two male donors (69 and 50 years old at the last data point) that span a time window of several years and that are divided into memory cells and plasmablasts, Fig. \ref{fig:data1}a. {Memory cells and plasmablasts were identified and sorted based on surface markers and their identity was verified using transcriptome lineage markers.} 
Each sample is divided into replicates (10X lanes) that are sequenced in parallel. 
The number of replicates per sample is 5 for memory cells and variable, from 1 to 7, for plasmablasts.

We assembled the \texttt{fastq} raw files using Cell Ranger (command vdj) \cite{zheng2017massively}, and then the V(D)J sequences are aligned and annotated using the Change-O toolkit \cite{gupta2015change}, which employs IgBLAST \cite{ye2013igblast} for the alignment to germline sequences provided by the IMGT database \cite{lefranc2009imgt}.
After these steps, we keep only cells that have a clean heavy and light chain pairing.
The size of samples divided into replicates are reported in Fig. \ref{fig:data1}b, with a total of around 30,000 cells per memory sample and a very variable number for plasmablasts (from a few hundred to few tens of thousands).
We excluded a sample from the original dataset for donor 1 taken in February 2021 because it contained both memory and plasmablast cells without the possibility of discrimination.
\rev{Clone clustering is performed with HILARy \cite{spisak2024combining} as described in the Methods section.}

\subsection{cDNA amplification {\em Dataset B}}

\begin{figure}[htbp]
	\centering
	\includegraphics[width=1\linewidth]{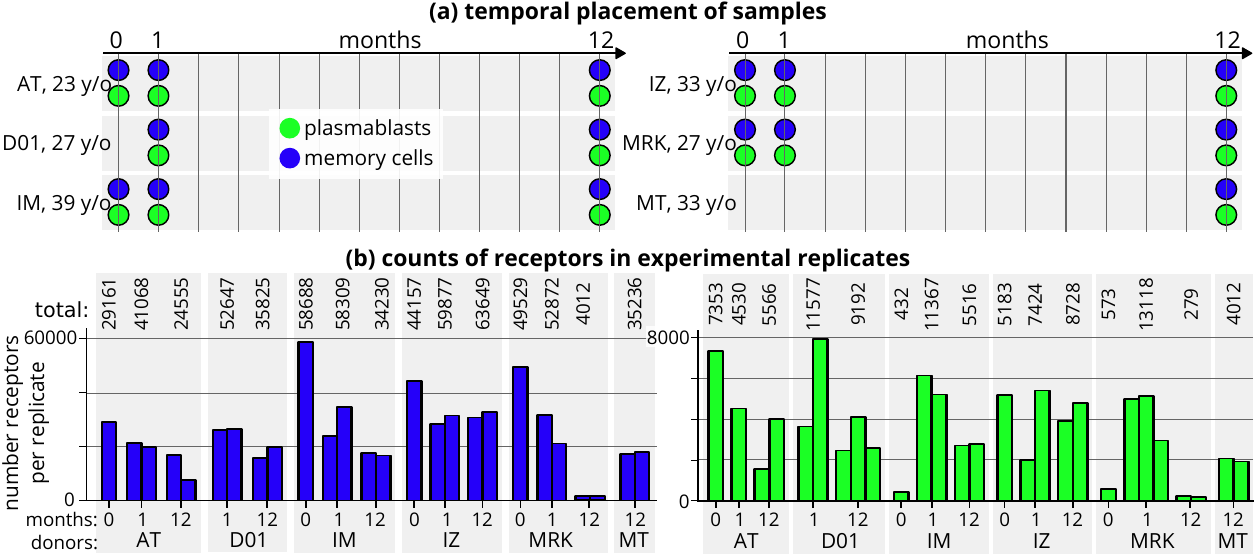}
	\caption{{\bf\em Dataset B} \cite{mikelov2022memory}
		(a): dataset samples of the six donors for memory cells (blue) and plasmablasts (green). 
		(b): counts of receptors for each experimental replicates for each sample. 
		On the left memory cell samples (blue), on the right plasmablasts (green).
	}
	\label{fig:data2}
\end{figure}

{\em Dataset B} \cite{mikelov2022memory} is from an IGH repertoire sequencing experiment with amplification of cDNA molecules (5'-RACE based protocol) using Unique Molecular Identifiers.
The dataset is available in the ArrayExpress repository under identifier \textit{E-MTAB-­11193}.
Using fluorescence-activated cell sorting (FACS), cells were sorted into memory cells and plasmablasts for six donors, aged 23 to 39 years old (Fig. \ref{fig:data2}a).
Samples are longitudinal for a maximum of 3 time points separated by 1 and 11 months, and can have up to three experimental replicates, see Fig. \ref{fig:data2}a.
For the donor MT, only the samples at the last time point are available.

We performed the raw data processing by using MiXCR \cite{bolotin2015mixcr} using the preset \texttt{mikelov-et-al-2021}.
For practical reasons, the obtained corrected sequences were also aligned with IgBLAST \cite{ye2013igblast} to germline sequences provided by the IMGT database \cite{lefranc2009imgt}.
The sizes of samples divided into replicates are reported in Fig. \ref{fig:data2}b, with a variable number of receptors: an average of 25,000 for memory cells and an average of $3500$ for plasmablasts.
As for the previous dataset, the clonal families were built with HILARy \cite{spisak2024combining}, with the difference that here we used the information only on the heavy chain.

\section{Likelihood computation and maximization for the noise model}
\label{sec:SM_ll_noise}

\subsection{Inference of the noise model with replicates}

Inference of the noise model is performed in each sample by exploiting its division into replicates.
We assume that the full repertoire of the donor is composed of $N_{\rm clone}$ different unique clones, each of them with a given count $n'_i$, $i=1, \ldots, N_{\rm clone}$.
By defining the total number of counts as $N_{\rm cell} = \sum_{i=1}^{N_{\rm clone}} n'_i$, we can define the corresponding frequencies as $f'_i = n'_i / N_{\rm cell}$.
As mentioned in the main text, we assume that the stationary distribution of these counts/frequencies is power-law distributed:
\begin{equation}
	f'_i \sim P_{\rm stat}(f' | {\bf p}_{\rm stat}) = \alpha (f_{\rm min})^\alpha f'^{-\alpha - 1} , \;\;\;\; f_{\rm min} \le f' \le 1 ,
	\label{eq:plaw}
\end{equation}
where we consider $f_{\rm min} = 1/N_{\rm cell} \ll 1$ and $\alpha > 1$ (consistent with observations).

This distribution depends on two parameters, ${\bf p}_{\rm stat} = (\alpha, f_{\rm min})$.
Notice that, overall, the repertoire frequencies are characterized by three parameters: $\alpha$, $f_{\rm min}$, and $N_{\rm clone}$. However, they will be reduced to $2$ since the frequency normalization imposes the constraint $\sum_i^{N_{\rm clone}} f'_i = 1$. 
For easier computations, we impose the average of this constraint, as also used in a previous work with a similar setting \cite{puelma2020inferring}:
\begin{equation}
	N_{\rm clone} \langle f' \rangle = 1 ,
	\label{eq:constraint}
\end{equation}
where the average is over the distribution $P_{\rm stat}$ in Eq. \ref{eq:plaw}.

A sequencing experiment consists of sampling these frequencies to obtain experimental counts of clones.
We call $\mathbf{n}^r = \{ {n}_1^r, {n}_2^r, \ldots \}$ the counts of clones in the replicate $r$, where ${N}_r$ is the total number of unique clones in the replicate.
The total number of cells in the replicate is ${M}_r = \sum_{i=1}^{N_r} {n}_i^r$.

We now want to write the probability of sampling ${n}_i^r$ cells of the clone $i$ in the replicate $r$, i.e., the noise model $P_{\rm noise}({n}_i^r | {\bf p}_{\rm noise})$.
In general, we want this probability to have an unbiased average $\mu_i^r = M f'_i$.
We consider two functional forms for the noise probability. 
The first is a Poisson distribution, which is completely defined by its mean and does not need any additional parameter.
The second is a negative binomial distribution parametrized by choosing the variance as ${\sigma^2}_{NB} = \mu + a \mu^b$.
We test a first case in which we fix $b = 1$, i.e., there is one free parameter ${\bf p}_{\rm noise} = a$, and variance is proportional to the average with a factor $1+a$.
For the second case we let the inference learn both the parameters ${\bf p}_{\rm noise} = (a, b)$.

\begin{figure}[htbp]
	\centering
	\includegraphics[width=0.85\linewidth]{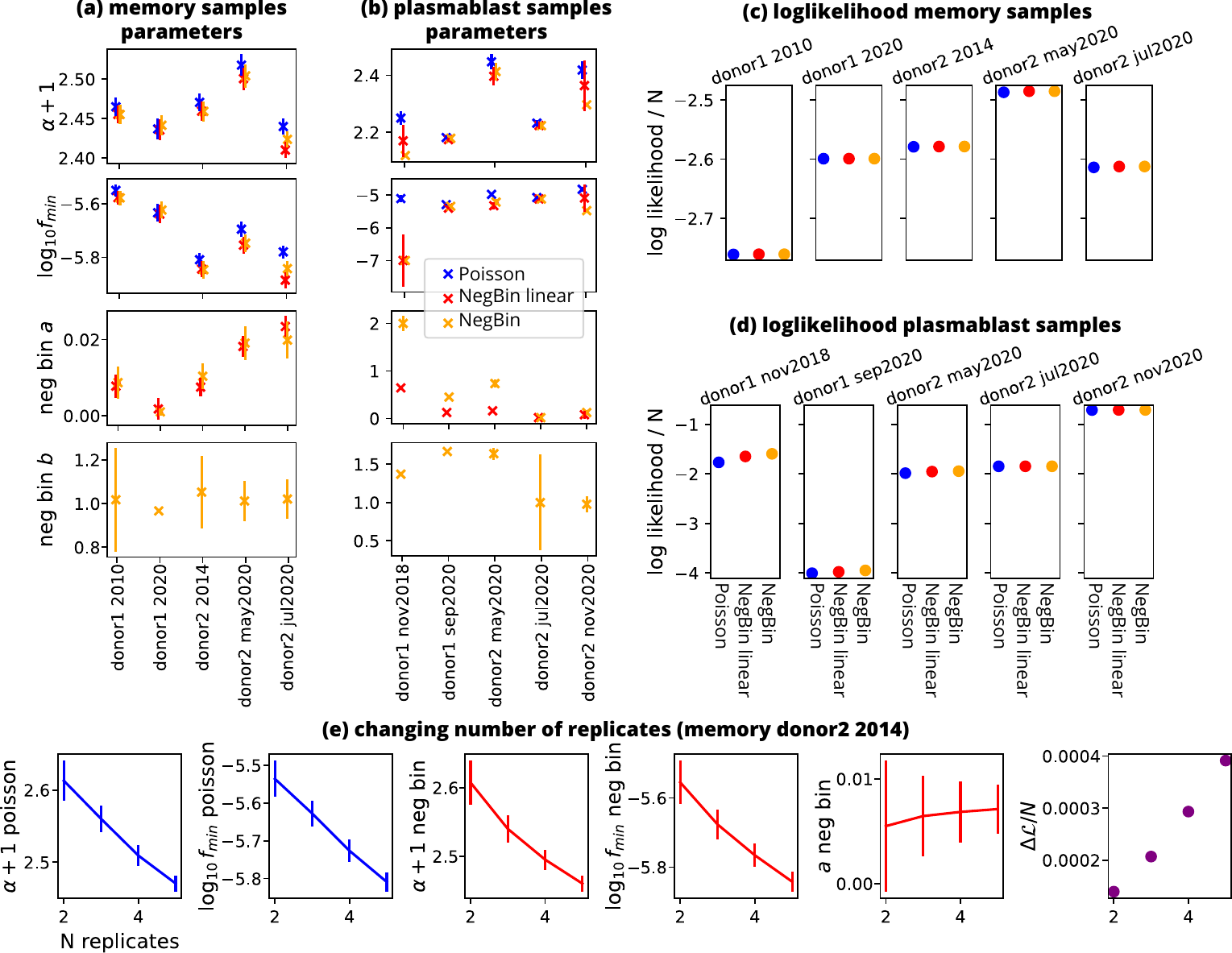}
	\caption{{\bf
	Results of the noise inference for {\em Dataset A}.}
	(a), (b): parameters learned in the three inference models represented with three different colors for samples of memory B-cells (a) and plasmablasts (b).
	The error bars are computed with the Hessian of the log-likelihood.
	(c), (d): values of the log-likelihood normalized by the number of clones for the three models (on the x-axis), for each sample, for memory B-cells (c) and plasmablasts (d).
	We run the noise inference of the Poisson and the linear Negative Binomial models by varying the number of replicates of donor 2, 2014 sample, \emph{Dataset A}.
	\rev{(e): dependence on the number of replicates of the inference (donor 2 2014, memory cells).
	For a given number of replicates ($2,3,4,5$) we run all the combinations among the 5 total replicates that are present. 
	We report the average learned parameter and its average error across all the combinations. 
	The two leftmost plots (in blue) show the Poisson parameters,
        the next three plots (in red) show the linear Negative Binomial parameters, and the rightmost plot is the difference of the log-likelihood per sample between the Negative Binomial and the Poisson model.}
	}
	\label{fig:infer_noise1}
\end{figure}

\begin{figure}[htbp]
	\centering
	\includegraphics[width=1\linewidth]{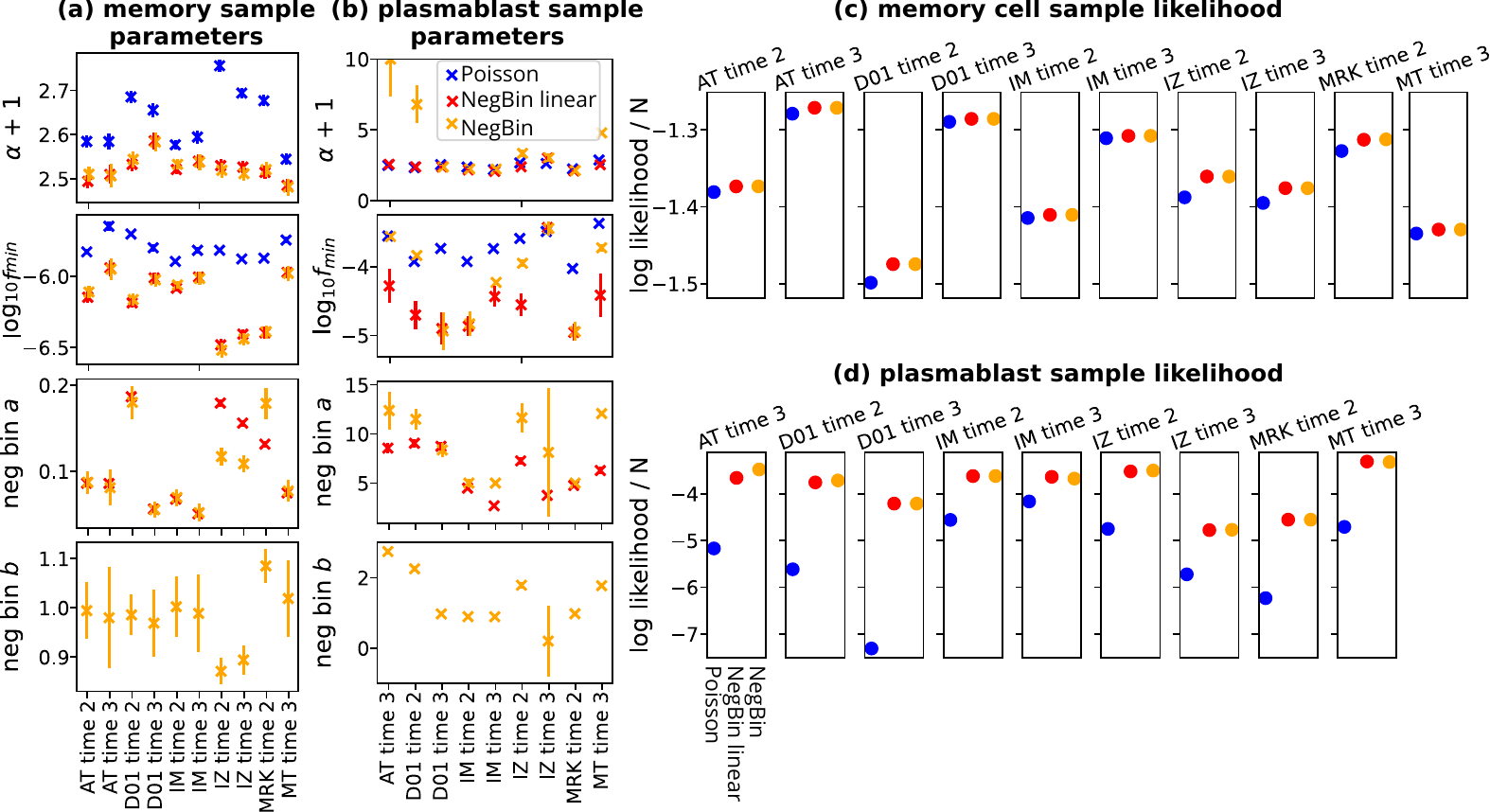}
	\caption{{\bf
	Results of the noise inference for {\em Dataset B}. }
	(a), (b): parameters learned in the three inference models represented with three different colors for samples of memory B-cells (a) and plasmablasts (b).
	The error bars are computed with the Hessian of the log-likelihood.
	(c), (d): values of the log-likelihood normalized by the number of clones for the three models (on the x-axis), for each sample, for memory B-cells (c) \rev{and} plasmablasts (d).
	}
	\label{fig:infer_noise2}
\end{figure}

\subsection{Likelihood maximization for the noise model}
\label{sec:ll_noise}

Given a dataset with $R$ replicates, $\{ \mathbf{n}^1, \ldots, \mathbf{n}^R \}$, we start by computing the probability that the cells of the clone $i$ have been observed with abundance $\{ {n}^1_i, \ldots, {n}_i^R \}$ in the replicates by using the power law distribution \eqref{eq:plaw} and one of the alternative noise probabilities discussed before:
\begin{equation}
	P({n}^1_i, \ldots, {n}_i^R | {\bf p}) = \int_{f_{\rm min}}^1 df' \; P_{\rm stat}(f' | {\bf p}_{\rm stat}) \prod_{r=1}^R P_{\rm noise}({n}_i^r | \mu = M_r f', {\bf p}_{\rm noise}) ,
	\label{eq:P_counts_clone_repl}
\end{equation}
where ${\bf p} = {\bf p}_{\rm stat} \cup {\bf p}_{\rm noise}$. That equation generalizes \SMRef{Eq. 4} to more than 2 replicates.

The next step is to compute the log-likelihood of observing the counts of the $N$ clones in the dataset, which is based on the assumption that they are independent:
\begin{equation}
	\mathcal{L}(\mathbf{n}^1, \ldots, \mathbf{n}^R | {\bf p}) = \sum_{i=1}^{N} \log P \left( {n}^1_i, \ldots, {n}_i^R \left| \sum_r {n}_i^r > 0, {\bf p} \right) \right. = \sum_{i=1}^{N} \log \left( \frac{P({n}^1_i, \ldots, {n}_i^R |{\bf p})}{1 - P(\mathbf{0} | {\bf p})} \right) .
	\label{eq:ll_noise}
\end{equation}
Notice that we conditioned the fact that a clone is present, i.e., $\sum_r n_i^r > 0$.
The second equality explicitly writes down this probability.

To get the best parameters that best describe the data, the final step is to maximize this likelihood over the set of parameters.
We have to remember that there is one constraint, Eq. \ref{eq:constraint}, that has to be satisfied during the maximization: 
\begin{equation}
	\theta^* = \argmax_{\theta} \mathcal{L}(\mathbf{n}^1, \ldots, \mathbf{n}^R | {\bf p})  \hspace{0.5cm} \text{satisfying} \hspace{0.5cm} \langle f' \rangle N_{\rm clone} = 1 .
	\label{eq:ML_noise}
\end{equation}

The results of this optimization are shown in Fig. \ref{fig:infer_noise1} and \ref{fig:infer_noise2} for the two datasets and for the three different noise models discussed before.
The panels \textit{a} and \textit{b} of the figure show the learned parameters ${\bf p}$: 2 for the stationary distribution, ${\bf p}_{\rm stat} = (\alpha, f_{\rm min})$, and the set of model-specific parameters ${\bf p}_{\rm noise}$ (none for the Poisson model, $a$ for the negative binomial with $b=1$, i.e., linear, and $(a,b)$ for the full negative binomial).
Panels \textit{c} and \textit{d} compare the log-likelihood per sample of the three models, allowing us to compare their performances and choose the most appropriate depending on the dataset.

\subsection{Numerical implementation of the likelihood maximization for the noise model}

The likelihood optimization is computed in Python, using wrapped functions in C++ to numerically solve the integrals for the probability of counts, Eq. \ref{eq:P_counts_clone_repl}.
The code can be found in the repository \url{https://github.com/statbiophys/memory_plasmab_dynamics}.

The integrals were solved with a standard trapezoid method using a logarithmic binning in the space of frequencies, with typically $10^4$ bins.
A useful trick to speed up the computation was to exclude from the integration all the frequency domain that was approximately zero for the integrand $g(f)$ of Eq. \ref{eq:P_counts_clone_repl}, since for many cases it is a peaked narrow function.
The threshold $\epsilon_0$ to determine if the integrand is approximately zero is chosen as the integrand maximum divided by $10^6$.
To move from $[f_{\rm min}, 1]$ to a new integration domain $[f_{\rm left}, f_{\rm right}]$ where the integrand is larger than the threshold, we scan the frequency axis by using a bisection algorithm that finds the values for which the integrand crosses the threshold $\epsilon_0$.

Optimization was performed in python using Brent's method implemented in the \textit{scipy} library, which allows for constrained optimization.
To speed up the computation, the summation terms of the likelihood in Eq. \ref{eq:ll_noise} are evaluated in parallel across multiple machine cores, using the \textit{multiprocessing} library of python.

\subsection{Errors estimates}
An estimate of the uncertainty of the optimal parameters can be obtained by the inverse of the Hessian of the log-likelihood, whose element corresponding to the parameters $p_i$ and $p_j$ is $H_{i,j} = - \partial_{p_i} \partial_{p_j} \mathcal{L}(\mathbf{n}^1, \ldots, \mathbf{n}^R | {\bf p})$.
In a likelihood maximization without constraints, the error associated with the parameter $p_i$ is $\sigma_i = \sqrt{H^{-1}_{i,i}}$.
However, we have to satisfy the constraint $C({\bf p}) = \langle f \rangle N - 1 = 0$.
To correctly include this in the uncertainty estimate, we need to project the Hessian on the hyperplane tangent to the constraint at the solution $p^*$.
This can be done by using the projection matrix $P = I - \mathbf{u}^{T} \mathbf{u}$, where $I$ is the identity matrix and $\mathbf{u} = \nabla C({\bf p}^*) / |\nabla C(\bf{p}^*)|$ is the normal unit vector to the constraint.
The local projection of the Hessian matrix along the constraint then reads $\hat{H} = P H P$.
The projected Hessian is no longer invertible since one eigenvalue is zero along the direction perpendicular to the constraint.
However, the error can be computed by summing all the contributions along each eigenvector of $\hat{H}$ and excluding that direction.
If we call $\lambda_k$ the eigenvalues of $\hat{H}$, putting the zero at $k=0$, i.e., $\lambda_0 = 0$, and $Q$ the matrix having the eigenvectors of $\hat{H}$ as columns, the error associated with the parameter $p_i$ is
\[
\sigma_i = \sqrt{\sum_{k \neq 0}  \frac{Q_{i,k} Q^{-1}_{k,i}}{\lambda_k} } .
\]

On a technical level, we realized that a numerical computation that evaluates $\mathcal{L}$ in ${\bf p}^*$, ${\bf p}^* - \delta {\bf p}$, ${\bf p}^* + \delta {\bf p}$ and uses the discrete formula for the second derivatives was leading to very unstable results, extremely sensitive to the size of $\delta {\bf p}$.
We decided then to compute analytically the first and second derivatives of Eq. \ref{eq:P_counts_clone_repl} with respect to all the parameters and evaluate the Hessian using those exact expressions.

\begin{figure}[htbp]
	\centering
	\includegraphics[width=0.95\linewidth]{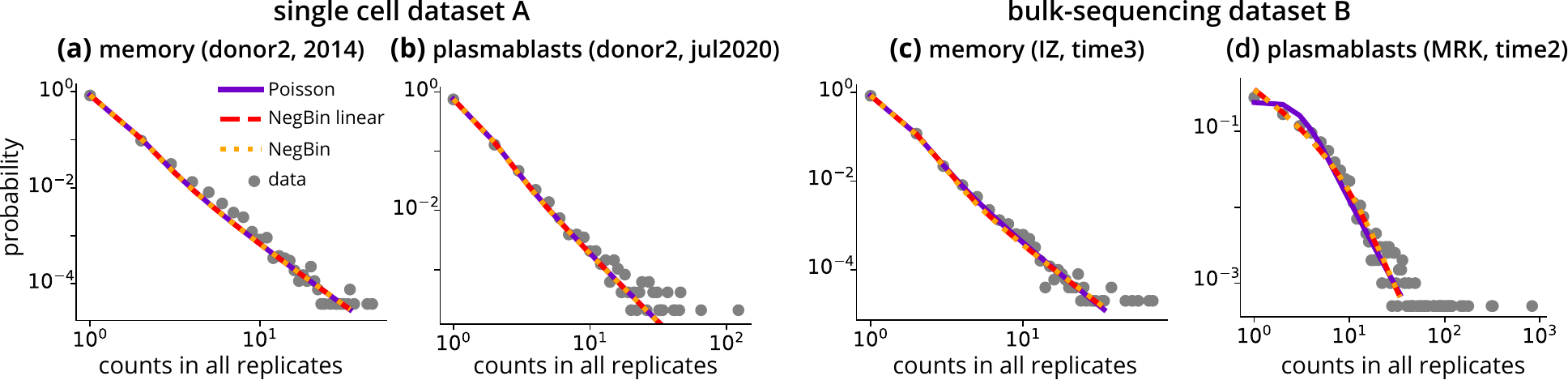}
	\caption{
	\textbf{Clone size distributions predicted by the noise model.}
		The empirical clone size distribution is compared with the prediction of the three inferred noise models: Poisson model, negative binomial with a fixed parameter $b=1$ and full negative binomial.
		We show the distribution of total counts in a sample $n$, which is the sum of the clone counts in each replicate: $n = \sum_r {n}^r$.
		(a): memory repertoire of donor 2 of {\em Dataset A},
                2014 timepoint, (b): plasmablast repertoire of donor
                2 of {\em Dataset A}, July 2020 timepoint, (c): memory
                repertoire of donor IZ of {\em Dataset B}, third time
                point), (d): plasmablast repertoire of donor MRK of
                {\em Dataset B}, second time point). 
	}
	\label{fig:noise_plaws}
\end{figure}

\section{Likelihood computation and maximization for the Geometric Brownian Motion}

\subsection{Geometric Brownian Motion}

The stochastic differential equation of the Geometric Brownian Motion (GBM) reads:
\begin{equation}
	\frac{d n'_i(t)}{dt}  = - \frac{n'_i}{\tau} + \frac{n'_i}{\sqrt{\theta}} \eta_i(t) ,
	\label{eq:SM_GBM}
\end{equation}
where $\tau$ is the time scale for the average exponential degradation, $\theta$ quantifies the amplitude of fluctuations, and $\eta_i(t)$ is a Gaussian white noise with the It\^o convention.
Clones go extinct when \rev{their abundance goes below $1$, which is mathematically modeled by introducing an} absorbing boundary at $n' = 1$, and new clones are introduced with a constant rate, $s$, starting with an initial abundance $n_0$.
This model is fully solvable by taking advantage of the substitution $x = \log n'$, which recovers a classical Brownian Motion for the log-size $x$.

The Fokker-Plank equation of the log-size reads:
\begin{equation*}
	\partial_t P(x,t) = \frac{1}{\tau}\partial_x P(x,t) + \frac{1}{2\theta} \partial^2_x P(x,t) + s \delta(x - x_0) ,
\end{equation*}
where $x_0 = \log n_0$.
This equation admits a stationary distribution that can be computed analytically:
\begin{equation}
	P_{\rm stat}(x) = \tau s e^{-\alpha x}
	\begin{cases}
		e^{\alpha x} - 1 & \text{if } x \le x_0
		\\
		e^{\alpha x_0} - 1 & \text{if } x > x_0
	\end{cases}
	\label{eq:SM_GBM_stat} ,
\end{equation}
where we introduce $\alpha = 2 \theta / \tau$.
This function shows a power law tail with exponent $-(\alpha + 1)$ in the limit of large abundance: $P_{\rm stat}(x) \sim e^{-\alpha x}$ and $P_{\rm stat}(n') = P_{\rm stat}(x)/n' \sim n'^{-\alpha - 1}$.
We should notice that Eq. \ref{eq:SM_GBM_stat} normalizes to the number of different clones in the repertoire $\int P_{\rm stat}(x) dx = N_{\rm clone}$, which allows us to find:
\begin{equation}
	N_{\rm clone} = \tau s \log n_0 = \frac{\alpha - 1}{\alpha} \frac{N_{\rm cell}}{n_0} \log n_0 ,
	\label{eq:SM_GBM_n_clones}
\end{equation}
where the second expression is rewritten by using Eq. \ref{eq:SM_GBM_n_cells} (see below) to remove the dependency on $\tau$, which is practically useful for estimating this number in the second dataset.
We can also find the total number of cells by solving the integral $N_{\rm cell} = \int n' P_{\rm stat}(n') dn'$:
\begin{equation}
	N_{\rm cell} = s (n_0 - 1) \left( \frac{1}{\tau} - \frac{1}{2 \theta} \right)^{-1}.
	\label{eq:SM_GBM_n_cells}
\end{equation}

Below we derive formulas to predict the survival properties of trajectories in a GBM used in the main text (\SMRef{Fig. 4}).
The first quantity to compute is the probability density of getting extinct, i.e. $x = 0$, after a time $t$ and starting from an initial condition $x_0$, the first passage time (FPT) density \cite{borodin2012handbook}:
\begin{equation*}
	p_{\rm FPT}(t | x_0) = x_0 \sqrt{\frac{\theta}{2 \pi t^3}} \exp \left( -\frac{\theta}{2 t} \left(x_0 - t/\tau \right)^2 \right) .
\end{equation*}
Closely related is the extinction probability at time $t$, which consists of summing up all the events that have led to hitting the boundary from $0$ to $t$:
\begin{equation*}
	p_{\rm extinct}(t | x_0) = \int_0^{t} dt'\;  p_{\rm FPT}(t' | x_0) = 1 - p_{\rm surv}(t | x_0),
\end{equation*}
where $p_{\rm surv}$ is the survival probability.
This allows us to compute, for example, the probability for a clone having abundance in $[x_{\rm min}, x_{\rm max}]$ to go extinct:
\begin{equation}
	f_{\rm turn}(t | x_{\rm min}, x_{\rm max}) = \frac{\int_{x_{\rm min}}^{x_{\rm max}} \; dx P_{\rm stat}(x) p_{\rm ext}(t | x)}{\int_{x_{\rm min}}^{x_{\rm max}} \; dx P_{\rm stat}(x) }.
	\label{eq:SM_turnover_frac}
\end{equation}
We can also interpret this quantity as the average fraction of trajectories that have been replaced by newly generated ones (in the stationary regime), i.e. the turnover fraction.

To move from the conditioning on $x$ to the experimental counts $n$, we write down:
\begin{equation*}
	p_{\rm FPT}(t, n) = \int \; dx \; P_{\rm stat}(x) \; P_{\rm noise}(n|M e^x / N_{\rm cell}) \; p_{FPT}(t | x) ,
\end{equation*}
and we can use it to compute the probability of a clone to get extinct at time $t$ given the experimental observation $n$ at the present time:
\begin{equation}
	p_{\rm surv}(t | n) = 1 - \frac{\int_0^t \; dt' p_{\rm FPT}(t',n)}{\int \; dx \; P_{\rm stat}(x) \; P_{\rm noise}(n|M e^x / N_{\rm cell})} .
	\label{eq:SM_p_surv_n_GBM}
\end{equation}

\subsection{Computing the likelihood with Monte Carlo sampling}

To write down the likelihood, we need the probability of observing the experimental counts of the clone $i$ in a given set of $T$ samples according to a model parametrized with ${\bf p}$:
\begin{equation}
	P(n^1_i, n^2_i, \ldots, n^T_i | {\bf p}) = \int \prod_{j=1}^T d n'^j P_{\rm dyn}(n'^1_i, n'^2_i, \ldots, n'^T_i | {\bf p}_{\rm dyn}) P_{\rm noise}(n^j_i | n'^j, {\bf p}_{\rm noise}) ,
	\label{eq:SM_pns}
\end{equation}
which generalizes \SMRef{Eq. 1} of the main text for an arbitrary number of samples.
The model, in this case a GBM, sets the joint probability over the hidden variables, $P_{\rm dyn}$, and the noise model, $P_{\rm noise}$, defines the probability of sampling an experimental count given the true hidden count.
If the analytical form of $P_{\rm dyn}$ is known, one can directly solve the integral with some numerical method like we did in Sec. \ref{sec:SM_ll_noise}.
The GBM admits an analytical solution for the propagator; however we choose an alternative approach to evaluate this integral through Monte Carlo (MC) integration.
A first reason for that is the fact that MC scales much better with the number of samples $T$ with respect to a multidimensional numerical integration.
Second, the next model that couples memory cells and plasmablasts is not analytically solvable but relatively easy to sample, making an MC-based approach the best choice for later.

The integral above can be evaluated with the classical expression of MC integration, where $N_{\rm MC}$ samples are generated from the dynamical model, and with $N_i^j$ we indicate the MC samples of experimental count for the clone $i$ at time $j$ sampled from the dynamical model, $\{N_i^1, \ldots N_i^T\} \sim P_{\rm dyn}(\cdot | {\bf p}_{\rm dyn})$:
\begin{equation}
	P(n^1_i, n^2_i, \ldots, n^T_i | {\bf p}) = \lim_{N_{\rm MC} \rightarrow \infty} \frac{1}{N_{\rm MC}} \sum_{N^1_i, \ldots, N^k_i \sim P_{\rm dyn}}^{N_{\rm MC}} \prod_{j=1}^T P_{\rm noise}(n^j_i | N_i^j, {\bf p}_{\rm noise}) .
	\label{eq:SM_MC}
\end{equation}
A practical problem of using this approach is that most of the sample generated by a GBM will have small real counts (the stationary distribution is dominated by small $n'$), leading to many samples for small experimental counts, i.e. $n = 0,1$, but very few samples for larger experimental counts, making the estimate of their probabilities very noisy.
A typical way to overcome the problem is to use importance sampling, where we introduce an arbitrary function $g(n'^1_i, n'^2_i, \ldots, n'^T_i)$ that samples more evenly our space.
The MC evaluation then becomes:
\begin{equation}
	P(n^1_i, n^2_i, \ldots, n^T_i | {\bf p}) = \lim_{N_{\rm MC} \rightarrow \infty} \frac{1}{N_{\rm MC}} \sum_{N^1_i, \ldots, N^T_i \sim g}^{N_{\rm MC}} \frac{P_{\rm dyn}(N^1_i, \ldots, N^T_i)}{g(N^1_i, \ldots, N^T_i)} \prod_{j=1}^T  P_{\rm noise}(n^j_i | N_i^j, {\bf p}_{\rm noise}) .
\end{equation}
We do not know, in general, the explicit shape of the function $P_{\rm dyn}$, making this expression not usable.
However, in general, we know the stationary distribution $P_{\rm stat}$, and we can split the joint probability as $P_{\rm dyn}(n'^1, \ldots, n'^T) = P_{\rm stat}(n'^1)P_{\rm prop}(n'^2, \ldots, n'^T | n'^1)$. 
The second object, $P_{\rm prop}$, is the propagator of the dynamical model that can be sampled by solving numerically the stochastic differential equation.
Similarly, we split the importance function into those two parts, with a first arbitrary function for sampling the initial point and the same propagator as before: $g(n'^1, \ldots, n'^T) = h(n'^1) P_{\rm prop}(n'^2, \ldots, n'^T | n'^1)$.
If we rewrite the equation above by plugging in these new definitions, we have:
\begin{equation}
	P(n^1_i, n^2_i, \ldots, n^T_i | {\bf p}) = \lim_{N_{\rm MC} \rightarrow \infty} \frac{1}{N_{\rm MC}} \sum_{N^1_i \sim h; N^2_i, \ldots, N^T_i \sim P_{\rm prop}}^{N_{\rm MC}} \frac{P_{\rm stat}(N^1_i)}{h(N^1_i)} \prod_{j=1}^T  P_{\rm noise}(n^j_i | N_i^j, {\bf p}_{\rm noise}) .
	\label{eq:SM_import_MC}
\end{equation}
In other words, the importance sampling is imposed only on the initial stationary distribution, $N^1_i \sim h$, for which we can explicitly compute the correction factor $P_{\rm stat}(N) / h(N)$, while the rest of the temporal samples are generated using the differential equation.
By choosing a proper shape of $h$, in our case, the variance of the likelihood can be reduced of more than one order of magnitude compared to Eq. \ref{eq:SM_MC}.

\subsection{Generating samples of the GBM}

The GBM is defined by 4 free parameters: $\tau$, $\theta$, $s$, and $n_0$.
We want to generate samples to compute Eq. \ref{eq:SM_import_MC} conditioned on the fact that the experimental counts at the first step are larger than zero: $P(n^1_i, n^2_i, \ldots, n^T_i | n^1_i > 0, {\bf p})$.
This condition allows us to exclude from the inference all the clones that are born between the first time point and the next ones, saving computational time in generating those events.

As a first step, we generate $N_{\rm MC}$ samples of log-counts at the first time point from the importance function $h(n^1)$.
This function is chosen as a gamma distribution over the space of the log-counts.
It depends on two parameters, the shape and the scale, which are chosen to minimize the likelihood variance.
This is done by selecting a set of GBM parameters {\bf p} reasonably close to the maximum and by computing many repetitions of the log-likelihood over which we compute the variance to minimize.
Each of the $N_{\rm MC}$ samples are then used as initial conditions of a numerical integration of Eq. \ref{eq:SM_GBM}.
The time step $dt$ is chosen to be smaller than all the time scales of the equation, i.e., $\tau$ and $\theta$, and typically of the order $0.1-0.01$ years. 
Log-counts that reach the boundary $x = 0$ are considered extinct.
After the desired interval of time, the log-count values are memorized and plugged into the noise probability $P_{\rm noise}$ to compute Eq. \ref{eq:SM_import_MC}.
The noise probability is dataset specific and independently learned previously, as described in Sec. \ref{sec:SM_ll_noise}.
Simulations of GBM trajectories are generated in C++ functions wrapped into Python.
The computation of Eq. \ref{sec:SM_ll_noise} given the generated samples, and the computation of the likelihood given those probabilities for each clone, are performed in Python.

\begin{figure}[htbp]
	\centering
	\includegraphics[width=0.85\linewidth]{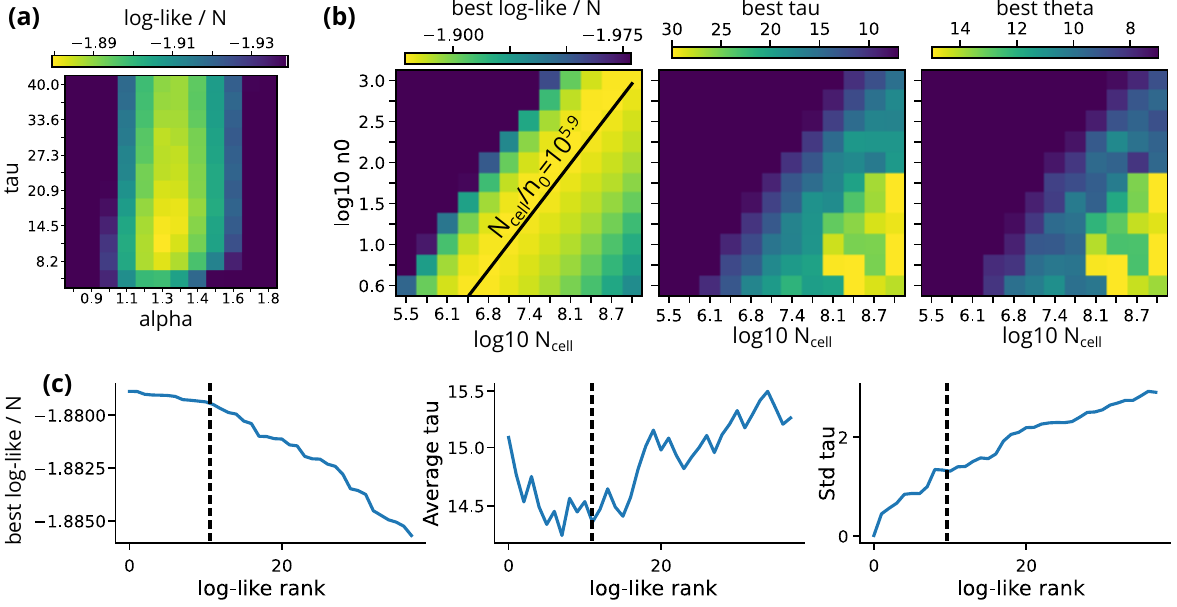}
	\caption{\textbf{Log-likelihood shape of the GBM for Dataset A.}
		(a): Log-likelihood computed for the first two time points of donor 2 from {\em Dataset A} at fixed $n_0=20$, $N_{\rm cell}=10^7$, and varying $\tau$ and $\alpha = 2 \theta / \tau$. 
		The Monte Carlo samples are $N_{\rm MC}=2 \cdot 10^6$.
		(b): Best likelihood for all three time points of donor 2, maximized over $\tau$ and $\theta$ computed at different values of $n_0$ and $M'$. The first plot is the likelihood value, the second plot is the $\tau$ that maximizes it and the third one is the best value of $\theta$.
		The black line fixes the ratio between $N_{\rm cell}$ and $n_0$ and passes at the maximum ridge.
		(c): Referring to the different inference procedures of panel (b) at different values of $N_{cell}$ and $n_0$, we rank the trials by likelihood. 
		We show, as a function of the best-likelihood rank, the best likelihood value, the average of best $\tau$'s across all the trials up to the rank on the x-axis, and the standard deviations of those $\tau$'s.
		The dashed black line is the chosen rank at which we can approximately say that the likelihood starts to drop. 
		The parameter values and errors for this donor are the average and standard deviations of all the inference trials up to this rank.
	}
	\label{fig:SM_GBM_loglike}
\end{figure}

\subsection{Optimizing the noisy likelihood with Gaussian Processes}

After the generation of MC samples,  and the computation of the probability of observing a sequence of experimental counts, the likelihood can be written down under the assumption of independent clones as:
\begin{equation}
	\mathcal{L}(\mathbf{n}^1, \ldots, \mathbf{n}^k | {\bf p}) = \sum_{i=1}^{N} \log P \left( {n}^1_i, \ldots, {n}_i^k \left| n_i^1 > 0, {\bf p} \right) \right. .
	\label{eq:ll_GBM}
\end{equation}
In general, we fix the number of Monte Carlo samples to $N_{\rm MC} = 3 \cdot 10^5$.
Since the probabilities are estimated from random samples, the likelihood itself is a noisy function that we want to maximize over the parameters ${\bf p}$.
We usually perform the optimization in the $\tau$-$\theta$ space or $\tau$-$\alpha$ space, fixing the values of the other two free parameters of the GBM, $s$ and $n_0$.
An efficient and reliable way to perform this stochastic optimization in a low-dimensional space is to use Gaussian Processes to approximate the likelihood. 
Each evaluation of the function reduces the uncertainty of its value, and the procedure tends to focus on the areas where it is more likely that the function has a maximum.
We used the function \textsc{gp\_optimize} of the \textsc{scikit-optimize} module of Python, \url{https://scikit-optimize.github.io/stable/index.html}.
There are several hyperparameters to fix. 
The kernel of the Gaussian Process is chosen as a \textit{Matérn kernel} plus a \textit{White kernel} that encodes prior knowledge on the amplitude of the MC fluctuations.
The \textit{Matérn kernel} depends on a length scale that sets the correlation lengths, imposing a prior on the smoothness of the function along the different axes.
In our case we have a very anisotropic function, Fig. \ref{fig:SM_GBM_loglike}a, which has a very long flat valley in one direction and a narrow section in the other.
This motivated us to use different length scales, $50-100$ times larger in the $\tau$ direction than in the $\alpha$ direction.
To balance the exploration-exploitation trade-off for sampling the function we use a \textit{negative expected improvement} acquisition.
We also set the \textit{improvement scale}, parameter \textsc{xi}, to be of the order of log-likelihood differences around the maximum, i.e., $2 \cdot 10^{-4}$.
A final hyper-parameter that we fix is the number of calls, \textsc{n\_calls}, around $200$, that is large enough to have a reliable estimate in a reasonable computational time.
As a result of the optimization, we consider the average minimum and not the minimal sample, which is actually the default result that the method returns.
Computationally, this can be done with the \textsc{skopt.expected\_minimum} function applied to the Gaussian process that has been learned at the previous step.

\begin{figure}[htbp]
	\centering
	\includegraphics[width=0.85\linewidth]{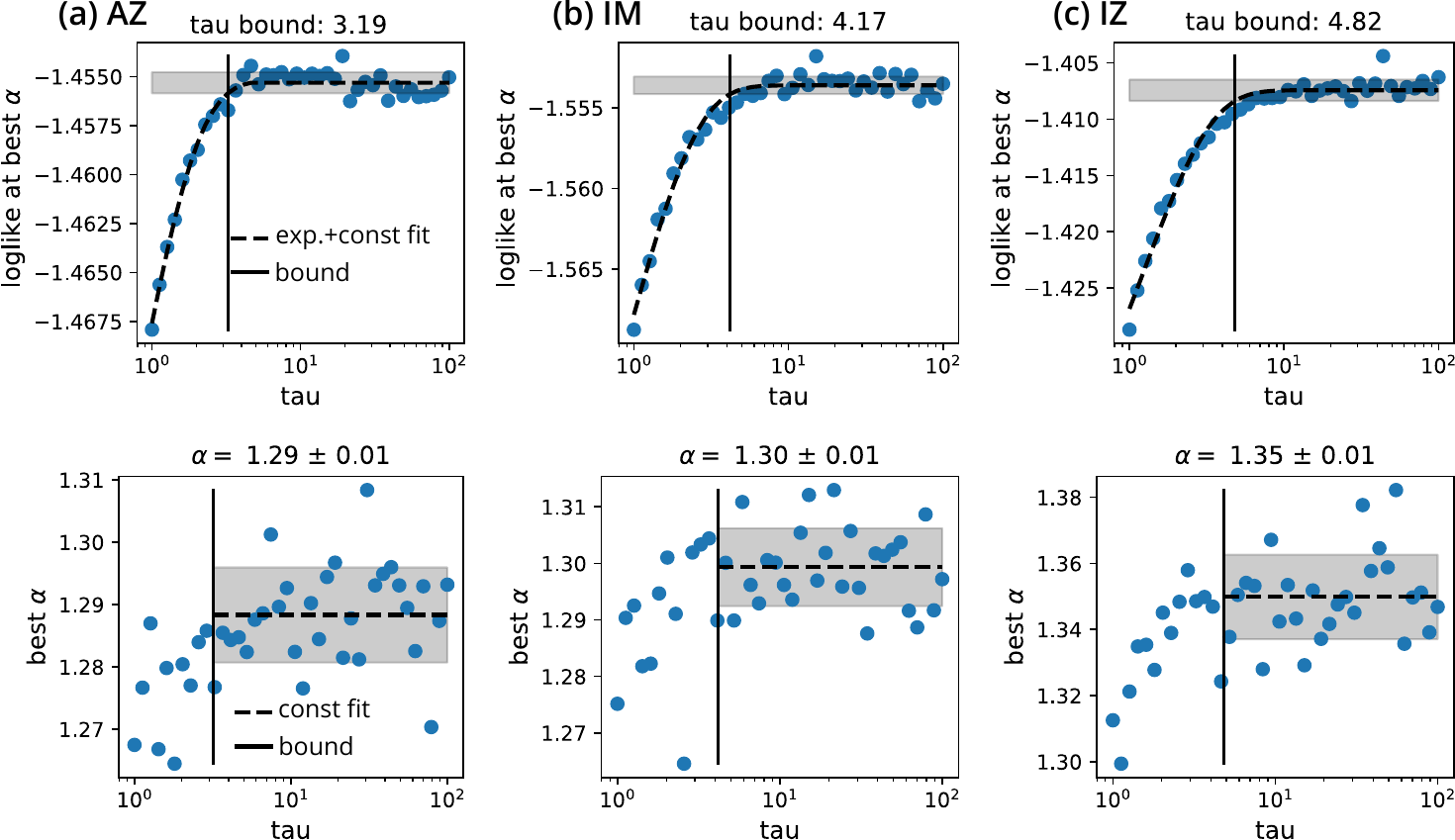}
	\caption{
	\textbf{Log-likelihood indeterminacy for the Dataset B.}	
	Log-likelihood shape as a function of $\tau_m$ for {\em Dataset B}.
	For the three donors with three time points, (a) AZ, (b) IM, (c) IZ, we show the log-likelihood that maximizes $\alpha$ at $N_{\rm cell}=10^8$ and $n_0=43$, $n_0=58$, $n_0=25$ respectively (each $n_0$ is chosen such that the $n_0/N_{\rm cell}$ ratio is optimal).
	In the upper panels we show the minus log-likelihood that optimizes over $\alpha$ at fixed $\tau$ for different values of $\tau$.
	The dashed line is a fit, as explained in the text. The shaded area shows the variability of points in the constant region, and the continuous line the position of the bound.
	The lower panels show, for the same inference procedure of the plot above, the best value of $\alpha$.
	Here the dashed line is a fit with a constant and the shaded area the variability in that region.
	}
	\label{fig:SM_GBM_Mikelov}
      \end{figure}

\subsection{Inference of the GBM parameters for the single-cell dataset ({\em Dataset A})}

The described procedure gives us the estimate of $\tau_m$ and $\theta_m$ for the memory cell repertoire at fixed values of two other free parameters of the GBM, $n_0$ and $s$. 
For practical reasons it is more convenient to consider $n_0$ and $N_{\rm cell}$ instead (using Eq.~\ref{eq:SM_GBM_n_cells}), since $s$ does not enter directly into the sampling procedure because of the conditioning $n^1 > 0$ that excludes the creation of new clones.
$N_{\rm cell}$ instead is used to evaluate the noise probability, and $n_0$ enters the stationary distribution expression.
We can run the inference for the two time points of donor 1 and for the 3 time points of donor 2 (results of \SMRef{Table 1}).
Notice that for \SMRef{Fig. 3f,g} of the main text we use the results from the first two time points to check how well the marginal prediction generalizes on the third time point.

Since, in principle, we do not know if the other two free parameters are learnable, we repeated the inference procedure several times at different values of $n_0$ and $N_{\rm cell}$, as shown in Fig. \ref{fig:SM_GBM_loglike}b.
The likelihood shows a ridge of constant altitude that is well fitted by a fixed ratio between $n_0$ and $N_{\rm cell}$, of order \rev{$N_{\rm cell}/n_0\sim 10^5 \textendash 10^6$}. 
\rev{More precisely, the ridge is located at $N_{\rm cell} / n_0 = (2.38 \pm 0.01) \cdot 10^5$ for the two time points of donor 1, and at $N_{\rm cell} / n_0 = (7.88 \pm 0.04) \cdot 10^5$ for the three time points of donor 2.
The fitting procedure considers a least square method where the points are $x = \log_{10} N_{\rm cell}$ and $y(x) = \max [\log_{10} n_0(x)]$ and the function is $y = x + c$, with $c = \log_{10}(n_0 / N_{\rm cell})$.
The error derives from the covariance matrix of the least square.}

The next question is how to choose the single value of $\tau$ and $\theta$ to associate to the donors among the different best parameters varying $n_0$ and $N_{\rm cell}$.
Fig. \ref{fig:SM_GBM_loglike}b shows that along the ridge those two parameters are not changing much.
To better account for this variability, we employ the following procedure.
We first rank the inference trials depending on the best likelihood value at different $n_0$ and $N_{\rm cell}$.
The first plot of Fig. \ref{fig:SM_GBM_loglike}c shows the best log-likelihood as a function of this rank, where we can approximately identify an initial plateau until rank $\sim 10$, which we can associate with the inference trials along the ridge.
We can also plot the average $\tau$ within the first top $r$ ranks as a function of $r$, second plot of Fig. \ref{fig:SM_GBM_loglike}c, and its standard deviation, third plot.
The average and standard deviation within the $\hat{r} = 10$ top ranks, i.e., along the ridge, will be our best estimate of the parameter and its uncertainty.
By varying the threshold rank $\hat{r}$, the average does not vary much, showing robustness against the arbitrary choice of $\hat{r}$.
In a similar way we can compute the estimate for $\theta$ or $\alpha$ and its uncertainty, \rev{which are shown in} \SMRef{Table 1}.

\rev{To compute $n_0$ and $N_{\rm clone}$, we use the estimate of the ratio $N_{\rm cell}/n_0$ and Eq. \ref{eq:SM_GBM_n_clones}, assuming the total number of cells to be in a range $N_{\rm cell} = 10^{10} \textendash 10^{11}$ (results in \SMRef{Table 1}). 
The birth rate $s$, Eq. \ref{eq:SM_GBM_n_cells}, depends on $N_{\rm cell}$ only through the ratio $N_{\rm cell}/n_0$.
Its uncertainty is computed by propagating the error on $\tau_m$, $\alpha$ and $N_{\rm cell}/n_0$.
}

\begin{figure}[htbp]
	\centering
	\includegraphics[width=\linewidth]{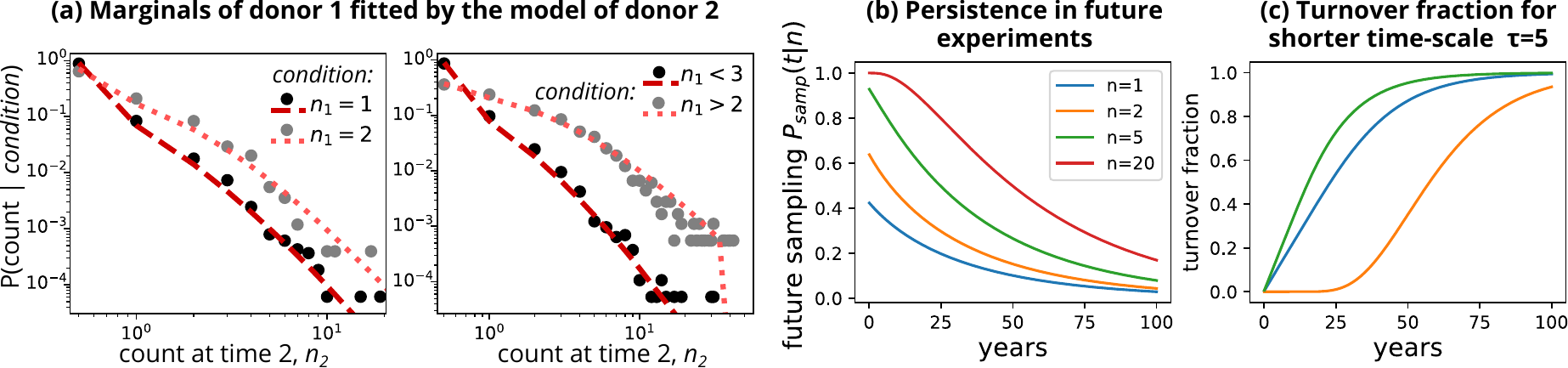}
	\caption{
		\textbf{\rev{Supplementary plots to the GBM inference.}}
		\rev{(a): Same plot of Fig \SMRef{3d,e} but with the data of donor 1 predicted by the model inferred from donor 2.}
		\rev{(b)}: Probability of sampling a clone in a future experiment after time $t$ having size $n$ in the present experiment.
		\rev{(c)}: Same as \SMRef{Fig. 3h} but for a shorter $\tau_m=5$ and $\theta_m = 3$.
	}
	\label{fig:SM_persist}
      \end{figure}

\subsection{Inference of the GBM parameters for the cDNA amplification dataset ({\em Dataset B})}

For the second dataset, we considered the three donors with sufficiently deep samples at all three time points: AZ, IM, and IZ.
Here we use the Negative Binomial with $b=1$ as a noise model, which depends on the parameter $a$.
This parameter is not learned for samples without replicates (the first time point of the three donors).
Since within a given donor values of $a$ are similar, for the sample without replicates we chose a value of $a$ that is the average of the other two values at any given donor.
This value is reported in \SMRef{Table 1} in the main text.

For the three donors, we found that the exponential decay time $\tau_m$ is not learnable, which can be seen in Fig. \ref{fig:SM_GBM_Mikelov}, upper panels.
The best likelihood's dependency over $\alpha$ for different values of $\tau_m$ does not show a maximum but it shows a constant plateau after a steady increase.
In other words, for $\tau_m$ large enough, including the case in which there is no decay at all, $\tau_m \rightarrow \infty$, we cannot see differences in the data.
As discussed in the main text, this is likely due to the small temporal resolution of the dataset, where the time points are separated by 1 year, while the decay time is probably of the order of $\sim 10$ years, and, in that small window of time, the slow variation is not noticeable.
However the value of $\alpha$ can be learned, because in the region of constant maximum likelihood, the best $\alpha$ is approximately constant (Fig. \ref{fig:SM_GBM_Mikelov} lower panels).

We can find a lower bound for $\tau_m$ by searching for the point at which the likelihood starts to flatten.
We fit the following piecewise function, i.e., an exponentially saturating function followed by a constant:
\begin{equation}
	\begin{cases}
		A - B e^{-C \tau_m} \;\; \text{for} \;\; \tau_m < \hat{\tau_m}
		\\
		A - B e^{-C \hat{\tau_m}} \;\; \text{for} \;\; \tau_m \ge \hat{\tau_m}
	\end{cases}
\end{equation}
The best fit is the dashed line in Fig. \ref{fig:SM_GBM_Mikelov}.
Then, we compute the standard deviation $\sigma$ of points in the region $\tau_m > \hat{\tau_m}$.
The intersection between the constant value $A - B e^{-C \hat{\tau_m}} - \sigma$ and the exponential for $\tau_m < \hat{\tau_m}$ gives us the lower bound for $\tau_m$, which is reported in \SMRef{Table 1} of the main text, and corresponds to the vertical line in the figure.

There is no minimum as a function of $\tau_m$, but we can still study the dependency on the parameters $n_0$ and $N_{\rm cell}$ as we did in Fig. \ref{fig:SM_GBM_loglike}b.
All three samples show a ridge compatible with a constant $n_0/N_{\rm cell}$ ratio, which we find to be \rev{$(2.11 \pm 0.03) \cdot 10^6$, $(1.70 \pm 0.01) \cdot 10^6$, and $(3.27 \pm 0.04) \cdot 10^6$} for donors AZ, IM, and IZ, respectively.
From those values and assuming \rev{$N_{\rm cell}=10^{10} \textendash 10^{11}$}, we can find the estimate of $n_0$ reported in \SMRef{Table 1} of the main text.
However, since we do not know the precise value of $\tau_m$, we can only find an upper bound for the introduction rate $s$.

\begin{figure}[htbp]
	\centering
	\includegraphics[width=0.5\linewidth]{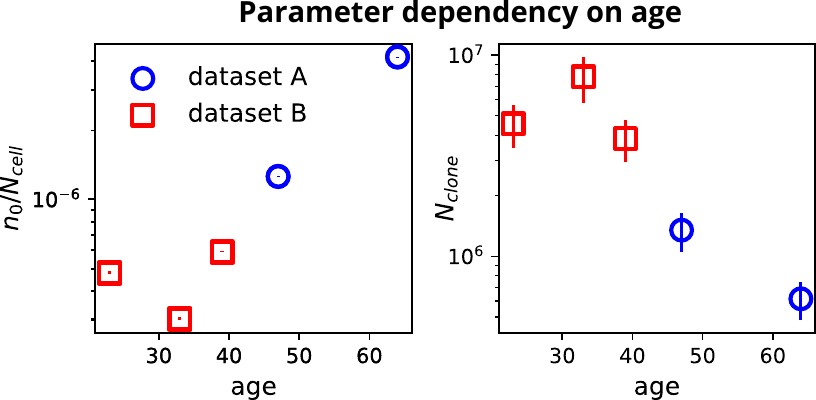}
	\caption{
		\textbf{Parameter dependency on age.}
		Normalized initial clone size, $n_0 / N_{\rm cell}$, and number of different clonal families $N_{\rm clone}$, versus age.
		\rev{Errors on the left plot are typically much smaller than the point.
			On the right plot the errorbars are the uncertainty depending on the estimate of $N_{\rm cell} \approx 10^{10} \textendash 10^{11}$.}
	}
	\label{fig:SM_age}
\end{figure}

\section{Likelihood computation for the plasmablast-memory dynamics}

\subsection{Coupled memory-plasmablast model}

The plasmablast dynamics assumes a Geometric Brownian Motion with an additional source term that models the differentiation of memory cells into plasmablasts.
Given a clone $i$, we label its number of memory cells in the body as $n_i'$ and its number of plasmablasts as $k_i'$, which follow the stochastic equation below:
\begin{equation}
	\left\{ 
	\begin{aligned}
		& \frac{d n'_i}{dt}  = - \frac{n'_i}{\tau_m} + \frac{n'_i}{\sqrt{\theta_m}} \eta_i(t) 
		\\
		& \frac{d k'_i}{dt}  = \rho n'_i - \frac{k'_i}{\tau_p} + \frac{k'_i}{\sqrt{\theta_p}} \xi_i(t) 
	\end{aligned}
\right. ,
\label{eq:SM_pb_dyn}
\end{equation}
where the first equation is Eq. \ref{eq:SM_GBM} for the memory cells and $\xi_i(t)$ is a a normal Gaussian noise.
The new set of parameters is $\rho$, the differentiation rate from memory to plasmablast, $\tau_p$, the plasmablast average decay time, and $\theta_p$, the time scale of the random fluctuations.
Unlike the simple GBM, the system is not analytically solvable anymore, but we can derive a useful relation between the average values of $n'$ and $k'$ at the stationary state:
\begin{equation}
	\langle k'_i \rangle = \rho \tau_p \langle n'_i \rangle, \hspace{1cm} K'_m = \rho \tau_p N_{\rm cell},
	\label{eq:SM_pb_stat}
\end{equation}
where the second equation is a consequence of the first one and connects the total number of plasmablasts generated by memory-cell differentiation, $K'_m$, to the number of memory cells, $N_{\rm cell}$.

Inferring all three plasmablast parameters from sampled repertoires is mathematically impossible.
This can be seen by writing the second equation for the frequencies: $f' = n'/N_{\rm cell}$ and $g' = m'/K'_m$, and using the relation between numbers of cells (Eq. ~\ref{eq:SM_pb_stat}):
\begin{equation}
	\frac{d {g'}_i(t)}{dt}  = \rho f'_i \frac{N_{\rm cell}}{K'_m} - \frac{{g'}_i}{\tau_p} + \frac{{g'}_i}{\sqrt{\theta_p}} \xi_i(t) = \frac{f'_i - {g'}_i}{\tau_p} + \frac{{g'}_i}{\sqrt{\theta_p}} \xi_i(t) .
\end{equation}
The dependency on $\rho$ disappears, and since experimental observations from samples depend only on frequencies and not on absolute counts, this parameter is not learnable.
To overcome this problem, we can use empirical estimates that connect the total number of the two groups of cells.
Define $K'$ as the total number of plasmablasts. $K'_m = \phi K'$, the number of memory-derived plasmablasts, is only a fraction $\phi$ of the total, since plasmablasts can also be generated by naive B cells.
We also consider the ratio between the total number of memory cells and the total number of plasmablasts $r = K' / N_{\rm cell}$.
The order of magnitude of these two numbers is known, and, in combination with Eq. \ref{eq:SM_pb_stat}, allows us to close the system of equations:
\begin{equation}
	K'_m = \phi r N_{\rm cell}, \;\; K'_m = \rho \tau_p N_{\rm cell}, \;\; \rightarrow \;\; \rho = \frac{\phi r}{\tau_p} .
	\label{eq:SM_pb_rho}
\end{equation}
Therefore, the strategy for the inference is to assume an empirically known value for $\phi$ and $r$ and to optimize the likelihood over $\tau_p$ and $\theta_p$, which set a value for $\rho$ through the equation above.

\subsection{Sampling the stochastic equation}

\begin{figure}
	\centering
	\includegraphics[width=0.8\linewidth]{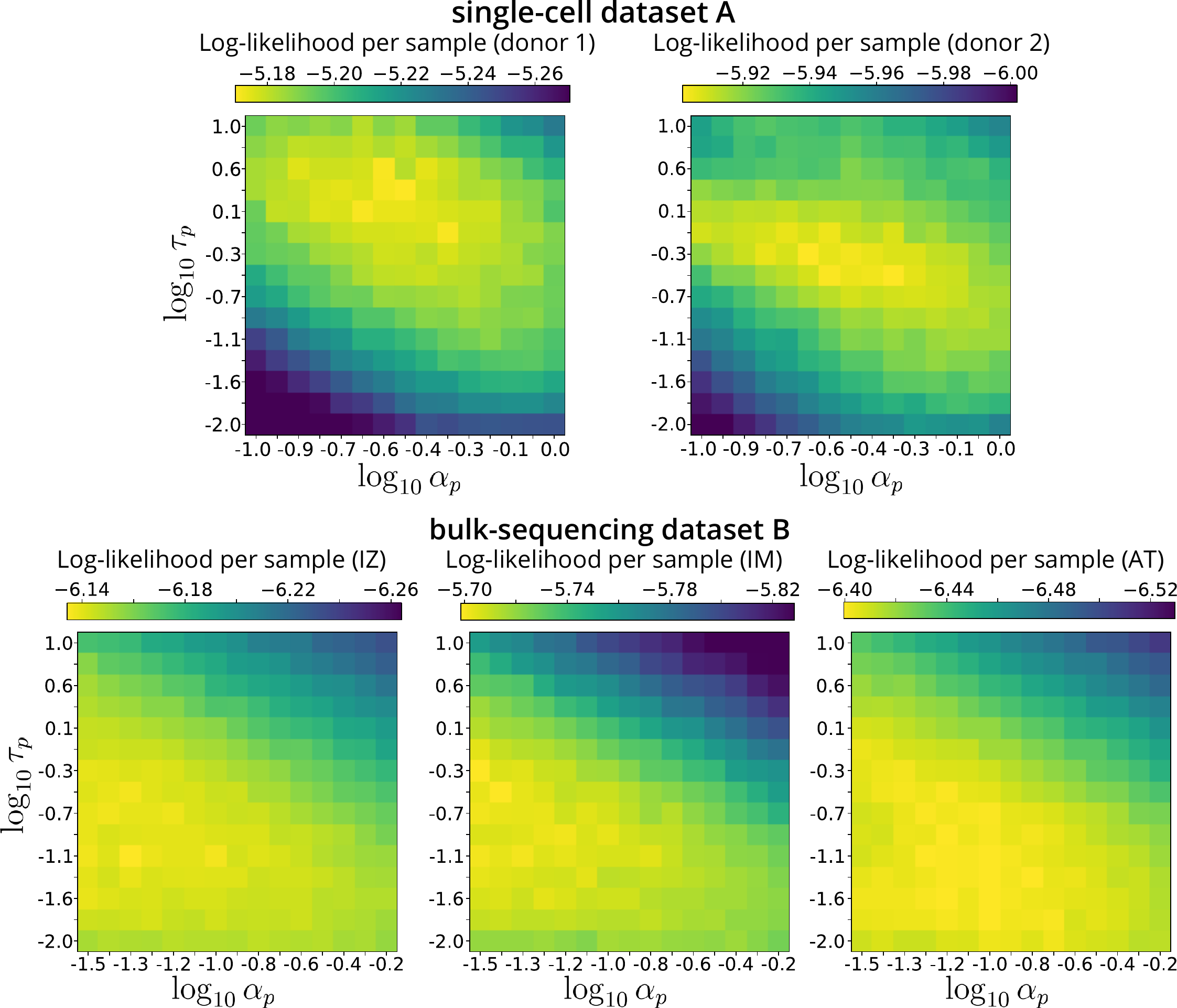}
	\caption{
		\textbf{Log-likelihood shape for the memory-plasmablast dynamics.}
		Log-likelihood computed by varying the two optimization parameters of the memory-plasmablast dynamics $\tau_p$ and $\alpha_p = 2 \theta_p / \tau_p$. 
		The number of Monte Carlo samples is $c \times N_{\rm
                  MC} = 5 \cdot 10^5$, where $c$ is the fraction of
                samples that exists at time 1 (see SI text).
		The parameters of the memory cell dynamics are fixed by the previous inference on memory cells. In the case of the second dataset for which we have only a lower bound on $\tau_m$, we chose $\tau_m = 5$ years, which is larger but close to the bound.
		The value of $n_0$ used in the simulations corresponds
                to the choice on $N_{\rm cell} = 10^{10}$.
		The plasmablast-memory ratio is chosen as $r = 0.1$, and the fraction of plasmablasts having a memory origin is $\phi = 0.5$.
		We show the log-likelihood for five experimental
                samples: donor 1 and 2 of {\em Dataset A} and three
                donors (IZ, IM, AT) of {\em Dataset B}. 
	}
	\label{fig:SM_pb_loglike}
      \end{figure}

To compute the likelihood, we need the probability of observing experimental counts given the parameters ${\bf p}_{\rm dyn} = (\tau_p, \theta_p)$, which has the shape of Eq. \ref{eq:SM_pns}.
Now we have some experimental samples of memory cells at different times $t$, $n_i^{t}$, and samples of plasmablasts $k^t_i$, and for each clone $i$, we want to compute $P(n^1_i, n^2_i, \ldots, k^1_i, k^2_i, \ldots | {\bf p})$.
Each sample has a noise model learned in \SMRef{Sec. 4A} of the main text, $P_{\rm noise}$, and the dynamical model $P_{\rm dyn}$, Eq. \ref{eq:SM_pb_dyn}, governs the dynamics of the real counts $n'$ and $k'$.
Notice that $P_{\rm noise}$ is defined by the average $\mu$, which is the frequency of the clone in the organism multiplied by total number of counts, which we call $M_p$ for plasmablasts.
Sequencing experiments contain all types of plasmablasts, not only the ones derived from memory cells, and therefore the frequency has to be computed by using $K'$: $\mu_i = M_p k'_i / K'$ for clone $i$.

An analytical solution for the propagator of Eq. \ref{eq:SM_pb_dyn} is not possible, and therefore we used the Monte Carlo integration to compute Eq. \ref{eq:SM_pns} as we did for memory samples.
We can also use the trick of importance sampling by choosing a proper distribution over the memory samples at the first time, $g$, and modify the summation of Monte Carlo samples as in Eq. \ref{eq:SM_import_MC}.
In the previous case we were conditioning on having the first experimental count $n^1_i > 0$, which was allowing us to neglect the dynamics of newly generated clones.
This condition, however, reduces the number of experimental samples, which is already not large for plasmablasts.
Therefore, we relax this constraint but we still impose the clone to be present in the memory samples, $\sum_j n^j_i > 0$, in such a way that we consider only plasmablasts that have been generated by memory cells.
As a consequence, in the MC simulation we need to generate new clones, which do not benefit from the importance sampling trick as before, since they are not sampled from the initial stationary distribution.
Putting these considerations together, we perform the Monte Carlo integration as a sum of two parts: one that estimates the probability conditioned on $n^1>0$ with importance sampling, and the second conditioned to $n^1 = 0$ without importance sampling.
The counterpart of Eq. \ref{eq:SM_import_MC} considers samples of memory-cell counts $N_i^t$ and of plasmablasts $K_i^t$, and it reads: 
\begin{equation}
	\begin{aligned}
	P(n^1_i, \ldots, k^1_i, \ldots | {\bf p}) = \lim_{N_{\rm MC} \rightarrow \infty} \frac{1}{N_{\rm MC}} \left[ \sum_{N^1_i \sim h; N^2_i, K^2_i \ldots \sim P_{\rm prop}}^{c N_{\rm MC}} \frac{P_{\rm stat}(N^1_i)}{h(N^1_i)} \prod_{j}  P_{\rm noise}(n^j_i | N_i^j, {{\bf p}}) \prod_{l}  P_{\rm noise}(k^l_i | K_i^j, {{\bf p}}) + \right.
	\\
	\left. + \sum_{N^1_i = 0; N^2_i,, K^2_i \ldots \sim P_{new}}^{(1-c) N_{\rm MC}} \prod_{j=1}^k  P_{\rm noise}(n^j_i | N_i^j, {{\bf p}}) \prod_{l}  P_{\rm noise}(k^l_i | K_i^j, {{\bf p}}) \right].
	\end{aligned}
	\label{eq:SM_import_MC_pb}
\end{equation}
The first summation samples $c\times N_{\rm MC}$ Monte Carlo clones (the fraction $c \le 1$ will be discussed later) that are present at the first step, using the importance trick, and we propagated them by simulating Eq. \ref{eq:SM_pb_dyn} with the initial condition for the plasmablasts given by Eq. \ref{eq:SM_pb_stat}.
The second summation considers non-existing clones at time $1$, $N^1_i = 0$, and it runs over the $(1-c)N_{\rm MC}$ generated new clones with a process labeled $P_{\rm new}$.
Each clone appears at a random time uniformly sampled in the interval between the first and last experiment (separated $\Delta t$).
Once created, the new clone has $n_0$ counts of memory cells and $1$ plasmablast, and from those initial conditions they follow Eq. \ref{eq:SM_pb_dyn}.
We checked that for different initial conditions on the number of plasmablasts the inference result does not change.
The fraction $c$ can be found by imposing that the number of different clones is constant in time and equal to the initial count $c N_{\rm MC}$.
By using Eq. \ref{eq:SM_GBM_n_clones}, we obtain the creation rate $s = c N_{\rm MC} / (\tau_m \log n_0)$, and the number of new clones is $s \Delta t = (1-c) N_{\rm MC}$, from which we get $c = 1 / [1 + \Delta t /(\tau_m \log n_0)]$.

The numerical simulation of Eq. \ref{eq:SM_pb_dyn} is performed as for the GBM case of memory cells.
One additional choice to make is when a plasmablast clone goes extinct, i.e., $k_i< 1$, but the memory cells of the same clone are not yet \rev{extinct}.
We choose to reintroduce the plasmablast at count $1$ after a time $t_{\rm new}$ given by the condition $\int_0^{t_{\rm new}} dt' n_i(t') \rho = 1$, that is, the time it takes on average for the memory cells to create 1 plasmablast.
One important technical point regards the choice of the discretization temporal step $dt$ in solving numerically the differential equation.
Unlike the previous case, here the time scales for the plasmablasts are smaller and we have to be careful of always choosing $dt$ smaller than the smallest time scale of the process.
This, in addition with a slightly more complex dynamics, makes the computational time of this second case slower than the previous one.

The likelihood maximization has been performed in the space $\log_{10} \tau_p$ and $\log_{10} \alpha_p = \log_{10} (2 \theta_p / \tau_p)$.
We transform the variables with a logarithm because we do not know {\em a priori} the orders of magnitude of these quantities.
The maximization was performed by using a Gaussian Process to smoothen the noisy landscape of the Monte Carlo estimated likelihood.
Unlike for memory cells, we do not impose different length-scale hyper-parameters in the kernel of the Gaussian Process.
We found that, when learnable, the log-likelihood landscape is not as anisotropic as in the memory cell case (see next Fig. \ref{fig:SM_pb_loglike}a).
We compute errors on the obtained optimal values by computing the Hessian of the average Gaussian Process function, which approximates the log-likelihood landscape in the absence of Monte Carlo noise. 
From the Hessian $H$, the error of parameter $p_i$ is $\sigma_i = \sqrt{H^{-1}_{i,i}}$.

\begin{figure}[htbp]
	\centering
	\includegraphics[width=0.8\linewidth]{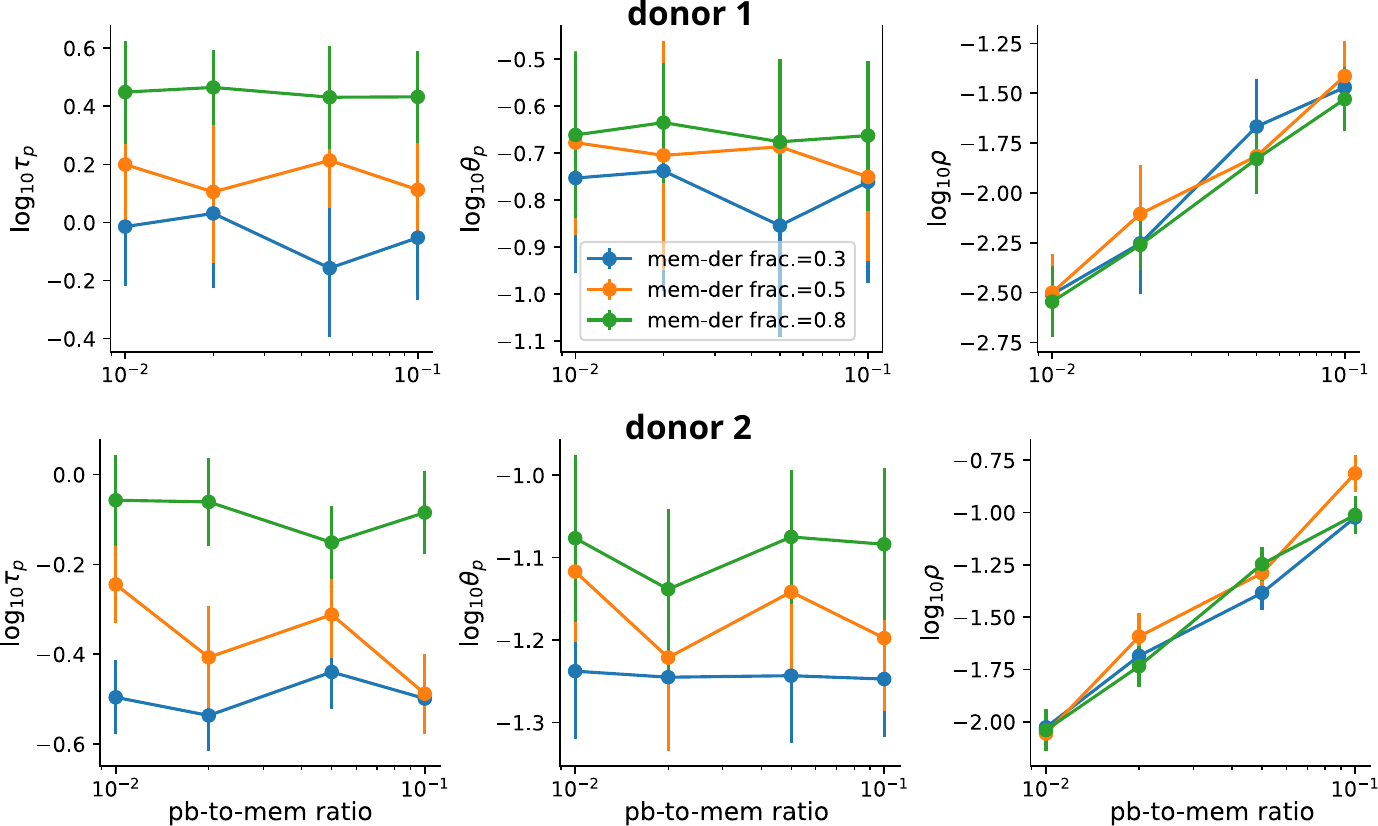}
	\caption{
		\textbf{Inferred parameters of the memory-plasmablast dynamics.}
		The parameters are computed
        across a range of different plasmablast-to-memory cell
        ratio, $r$, on the x-axis, for different values of the
        fraction of plasmablasts that derive from memory
        cells, $\phi$ (colors).
		The error bars are computed using the inverse of the likelihood Hessian, where the Hessian is obtained from the smoothed approximation of the likelihood by the Gaussian Process that is used to maximize the noisy function.
	}
	\label{fig:SM_pb_params}
\end{figure}

\subsection{Likelihood optimization for {\em Dataset A}}

We perform the inference procedure for both donors.
For the first donor we exclude the first plasmablast sample given its small depth and we use the remaining two memory and two plasmablast samples.
For the second donor we run the inference by using all three samples of memory cells and the second and third ones of plasmablasts, excluding the first sample for reasons of sequencing depth, and the last sample for reasons of performance.
However, we checked that including it does not change the results.
The inference uses the memory cell parameters learned at the previous step and $n_0$ fixed by the choice $N_{\rm cell}=10^{10}$.
This leads to the likelihood landscape shown in Fig. \ref{fig:SM_pb_loglike}. It shows a clear maximum, which still remains when varying the empirical parameters $r$ and $\phi$.
The results for different learning trials at different $r = 0.01 \textendash 0.1$ and $\phi = 0.3 \textendash 0.8$ are shown in Fig. \ref{fig:SM_pb_params}, where $\tau_p$ and $\theta_p$ are learned and $\rho$ is obtained from the constraint \ref{eq:SM_pb_rho}.

\subsection{Likelihood optimization for {\em Dataset B}}

The longitudinal structure of this dataset, with relatively short intervals between timepoints, makes it suitable to learn the fast plasmablast dynamics.
However, we obtain a flat likelihood surface that keeps going for increasingly smaller values of $\alpha_p$.
This flatness does not allow us to learn the parameters. We suspect that the reason is the noise in the experimental plasmablast samples, for which we have learned a Negative Binomial model with a much larger variance than a Poisson model (see Sec. \ref{sec:SM_ll_noise}). 
The average decay of plasmablasts and the relation with memory cells is hidden in the noise. Consistently, the best likelihood area includes region of maximal noise $\alpha_p \rightarrow 0$, i.e., $\theta_p \rightarrow 0$.


\begin{figure}[htbp]
	\centering
	\includegraphics[width=\linewidth]{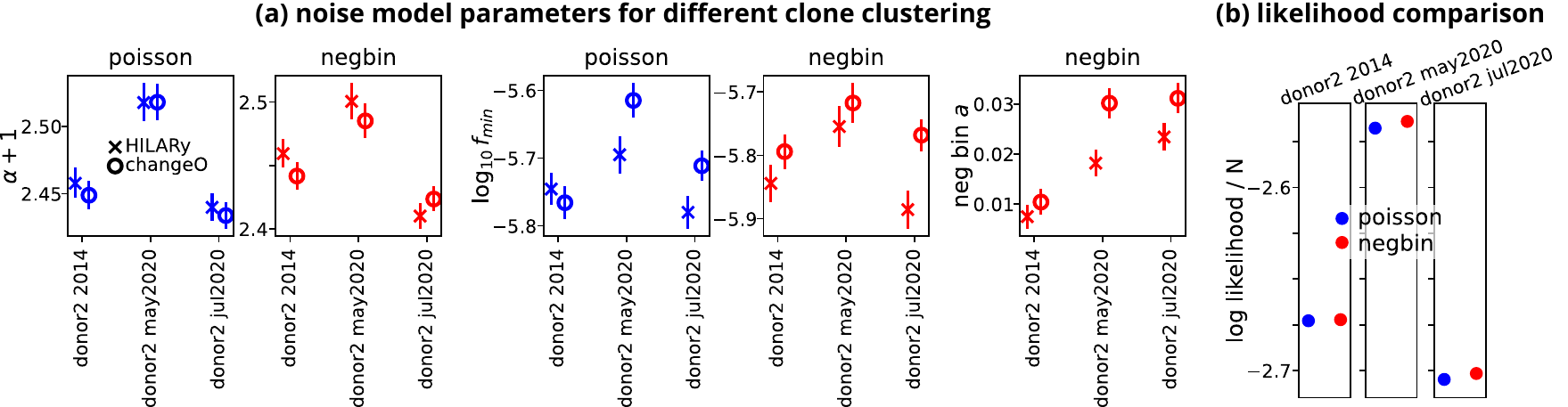}
	\caption{
		\rev{
			\textbf{Comparison of the noise parameters
                          inferred using HILARy and ChangeO inferred families.}
			Poisson (blue) and Negative binomial (red)
                        model parameters inferred from the memory
                        B-cell samples grouped into clones using two
                        different clustering algorithms, HILARy and
                        ChangeO (shown with different point markers), for the second donor of \emph{Dataset A}.
			(a) model parameters computed as described in Sec. \ref{sec:SM_ll_noise}.
			(b) log-likelihood per sample for the ChangeO families.
		}
	}
	\label{fig:SM_noise_changeO}
\end{figure}

\begin{table*}[htbp]
	\begin{tabular}{c||c|c|c|c|c|c|c|c|c} 
		& decay time, & fluct. time & power law & initial clone & rate of new & number of & diff. rate & decay time & fluctuation\\
		& $\tau_m$ (year) & $\theta_m$ (year) & exponent, $\alpha$ & size, $n_0$ & clones, $s$ (1/year) & clones, $N_{\rm clone}$ & $\rho$ (1/year) & $\tau_p$ (year) & time $\theta_p$ (year) \\ [0.5ex] 
		\hline \rule{0pt}{1\normalbaselineskip}
		Hilary & $14.5 \pm 1.3$ & $8.7 \pm 0.7$ & $1.20 \pm 0.04$ & $10^{4} \textendash 10^5$ & $(9.0 \pm 1.6) \cdot 10^3$ & $(1.2 \textendash 1.5) \cdot 10^6$  & $0.01\textendash0.1$ & $0.25\textendash1$ & $0.05\textendash0.1$ \\
		ChangeO & $13.0 \pm 1.0$ & $7.6 \pm 0.4$ & $1.17 \pm 0.04$ & $10^{4} \textendash 10^5$ & $(8.3 \pm 1.7) \cdot 10^3$ & $(1.0 \textendash 1.3) \cdot 10^6$ & $0.01\textendash0.1$ & $0.3\textendash1$ & $0.05\textendash0.09$ \\
	\end{tabular}
	\caption{
		\rev{
		\textbf{Comparison of the parameters of the dynamical models
                          inferred using HILARy and ChangeO inferred families.}
		Results are presented for donor 2 of \emph{Dataset A}
                for both the memory B cell and plasmablasts models. 
		Note that the first line (HILARy) corresponds to the
                numbers reported in \SMRef{Tables 1 and 2}.
	}
	}
	\label{tab:robust}
\end{table*}

\begin{table*}[htbp]
	\centering
	\begin{tabular}{c||c|c} 
		& decay time, & fluctuation \\
		& $\tau_m$, years & time, $\theta_m$ \\ [0.5ex] 
		\hline \rule{0pt}{1\normalbaselineskip}
		donor 1 (A) & $8.8 \pm 1.8$ & $6.5 \pm 1.6$ \\ 
		donor 2 (A) & $9.6 \pm 2.4$ & $6.5 \pm 1.5$ \\ [1ex] 
	\end{tabular}
	\caption{
		Inferred parameters of the Geometric Brownian Motion
                for the memory B-cell dynamics obtained by defining a
                clone as a set of identical sequences, for donor 2 of \emph{Dataset A}.}
	\label{tab:SM_params_gbm_sameseq}
\end{table*}

\begin{figure}[htbp]
	\centering
	\includegraphics[width=0.5\linewidth]{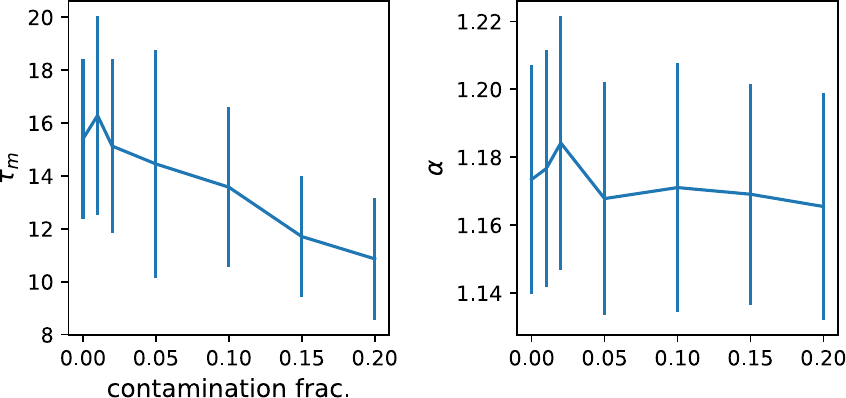}
	\caption{
		\rev{
			\textbf{Model inference with contamination of plasmablasts in the memory samples.}
		We repeated the memory-cell GMB inference by
                substituting a fraction (x axis) of memory cells with
                plasmablasts taken at the same time, for donor 2 of \emph{Dataset A}.
			The inference fixes $N_{\rm cell} = 10^{10}$ and $n_0=12700$ (a point along the ridge of maximum likelihood of Fig. \ref{fig:SM_GBM_loglike}) and finds the best $\tau_m$ and $\alpha$ by using the Gaussian Process method described in Sec. \SMRef{S3}.
			The error is computed with the inverse Hessian of the log-likelihood approximated with the Gaussian process.
			Notice that, for fraction 0, the precise
                        values and errors are slightly different than
                        in \SMRef{Tab. 1} because the reported
                        parameters are averaged along the $N_{\rm cell}-n_0$ ridge as explained in Sec \SMRef{S3}.
		}
	}
	\label{fig:SM_contamination}
      \end{figure}

\end{document}